# X-ray Constraints on the Intrinsic Shape of the Lenticular Galaxy NGC 1332

David A. Buote[1] and Claude R. Canizares[2]

Department of Physics and Center for Space Research 37-241,

Massachusetts Institute of Technology

77 Massachusetts Avenue, Cambridge, MA 02139

## ABSTRACT

We have analyzed ROSAT PSPC X-ray data of the optically elongated S0 galaxy NGC 1332 with the purposes of constraining the intrinsic shape of its underlying mass and presenting a detailed investigation of the uncertainties resulting from the assumptions underlying this type of analysis. The X-ray isophotes are elongated with ellipticity $0.10 - 0.27$ (90% confidence) for semi-major axes $75'' - 90''$ and have orientations consistent with the optical isophotes (ellipticity $\sim 0.43$). The spectrum is poorly constrained by the PSPC data and is consistent with a single-temperature Raymond-Smith plasma ($T \sim 0.6$ keV). However, the spectrum cannot rule out sizeable radial temperature gradients $\left|\frac{d \ln T_{gas}}{d \ln R}\right| < 0.35$ (99% confidence) or an emission component due to discrete sources equal in magnitude to the hot gas. Using (and clarifying) the geometric test for dark matter introduced by Buote & Canizares (1994), we determined that the hypothesis that mass-traces-light is not consistent with the X-ray data at 68% confidence and marginally consistent at 90% confidence independent of the temperature profile of the gas; similar results are obtained considering the effects of possible gas rotation and emission from discrete sources. Detailed analysis of the mass distribution following Buote & Canizares gives constraints on the ellipticity of the underlying mass of $\epsilon_{mass} = 0.47 - 0.72(0.31 - 0.83)$ at 68% (90%) confidence for isothermal and polytropic models. The total mass of the isothermal models within $a = 43.6$ kpc ($D = 20h_{80}^{-1}$ Mpc) is $M_{tot} = (0.38 - 1.7) \times 10^{12} M_\odot$ (90% confidence) corresponding to total blue mass-to-light ratio $\Upsilon_B = (31.9 - 143)\Upsilon_\odot$; polytropic models yield mass ranges larger by a factor of $\sim 2$ due to the uncertainty in the temperature profile. Similar results are obtained when the dark matter is fit directly using the known distributions of the stars and gas. When possible rotation of the gas and emission from discrete sources are included flattened mass distributions are still required, although the constraints on $\epsilon_{mass}$, but not the total mass, are substantially weakened.

---

[1] dbuote@space.mit.edu

[2] crc@space.mit.edu

*Subject headings:* dark matter — galaxies: elliptical and lenticular, cD — galaxies: individual (NGC 1332) — galaxies: photometry — galaxies: structure — X-rays: galaxies

## 1. Introduction

The distribution of intrinsic shapes of galactic halos is of cosmological importance since it can be predicted from cosmological N-body simulations. Cold Dark Matter (CDM) simulations predict dark halos that are flatter ($\epsilon \sim 0.5$) than inferred from analysis of observed optical isophotes ($\epsilon \sim 0.3$) of ellipticals (e.g., Dubinski & Carlberg 1991; Dubinski 1994); the distribution of the shapes of halos in CDM simulations appears to be insensitive to the power spectrum (Frenk et al. 1988; Efstathiou et al. 1988). Since the observed stars may not trace the shape of the dark mass, this is not necessarily a discrepancy: the shape of the total gravitating matter is required to compare to the simulations. Unfortunately, reliable constraints on the shapes of dark halos exist for only a few galaxies (Sackett et al. 1994). The uncertainty in the shape of the stellar velocity dispersion tensor has hindered optical methods to measure the dark matter distribution in early-type galaxies. In fact, the need for any dark matter in early-type galaxies has not been definitively established from optical data (Kent 1990; de Zeeuw & Franx 1991; Ashman 1992; although see Saglia et al. 1993). Measuring the intrinsic shapes of early-type galaxies using the observed velocity profiles may eventually enable robust constraints from optical data (Statler 1994).

Perhaps the most powerful probe of the intrinsic shapes of early-type galaxies is the hot, X-ray–emitting gas because the dispersion tensor of the gas should be isotropic (e.g., Sarazin 1986). The full potential of X-ray analysis for determining the shapes of dark halos has not yet been achieved primarily because instruments on board previous X-ray satellites lacked the spatial resolution to accurately measure the shapes of the X-ray isophotes (White & Canizares 1987; White 1987). Recently, Buote & Canizares (1994; hereafter BC94) used the higher resolution X-ray data of the ROSAT PSPC ($FWHM \sim 30''$) to constrain the shape of the E4 galaxy NGC 720. By assuming the X-rays are due to hot gas in hydrostatic equilibrium with the galactic potential, BC94 obtained for the halo ellipticity, $\epsilon = 0.46 - 0.74$ (90% confidence), which is mostly insensitive to the poorly determined temperature profile of the gas. In addition, BC94 introduced an X-ray test for dark matter in early-type galaxies that is completely independent of the temperature profile of the gas and, in principle, without requiring any model fitting; we elucidate the robust nature of the test (§5.1). Moreover, this geometric test may be applied to test alternate gravity theories like MOND.

This "geometric" test is especially relevant considering that the lack of high quality, spatially resolved temperature profiles for most early-type galaxies has rendered inconclusive previous X-ray studies of dark matter (e.g., Fabbiano 1989). Even for those galaxies now possessing high quality, spatially resolved spectra from ROSAT, the translation of these spectra into temperature profiles using standard spectral models is still uncertain (e.g., Trinchieri et al. 1994). This uncertainty in interpreting temperatures may be resolved with X-ray data from ASCA (Tanaka et al. 1994). Although the poor spatial resolution of ASCA inhibits measurements of temperature profiles, the superior spectral resolution allows for model-independent determination of mean galactic temperatures from line ratios that provides preliminary evidence for dark matter in three bright ellipticals in Virgo (Awaki et al. 1994). Since establishing the firm existence of dark matter in early-type galaxies is of vital consequence to cosmological theories of structure formation (e.g., Ashman 1992; Silk & Wyse 1993), an X-ray method to detect dark matter that takes advantage of the high quality spatial data of current X-ray satellites but does not require detailed temperature information is of great utility; note that methods to map the projected mass density of individual galaxies with weak gravitational lensing (Kaiser & Squires 1993) are impractical because the angle subtended by galactic halos at the redshifts required for significant lensing effect contains too few background galaxies for an adequate signal; however, it may be possible to usefully probe the outer portions of galactic halos with this method (i.e. $r \gtrsim 100$ kpc).

In this paper we analyze ROSAT PSPC data for the S0 galaxy NGC 1332 which, along with NGC 720, is especially suited for X-ray analysis of its intrinsic shape. NGC 1332 is quite elongated in the optical ($\epsilon \sim 0.43$), a desirable property to reduce the likelihood of substantial projection effects. If, in addition, the elongation in the optical is correlated to intrinsic elongation of the underlying mass, then the gravitational potential, hence the X-ray isophotes, should exhibit noticeable flattening. Located in a poor cluster (the Eridanus group, Willmer et al. 1989), NGC 1332 is relatively isolated from other large galaxies. Isolation is desirable so that the gas is not distorted by ram-pressure or tidal effects (Schechter 1987). The X-ray emission of NGC 1332 extends to over $7'$ on the sky, thus providing many pixels of angular resolution. Considering the desired elongation, isolation, and angular sizes, NGC 1332, along with NGC 720, possess the largest X-ray fluxes of normal early-type galaxies observed with *Einstein* (Fabbiano, Kim, & Trinchieri 1992) that are likely to be dominated by emission from hot gas (Kim, Fabbiano, & Trinchieri 1992).

We explore in detail in this paper many of the fundamental assumptions underlying X-ray analysis of the intrinsic shapes and radial mass distributions in early-type galaxies to provide a reference for future investigators. These issues were discussed in our study of NGC 720 (BC94), but the peculiarities of that galaxy did not require detailed investigation of these issues (e.g., position-angle offset of the X-rays and optical light in NGC 720 argues against substantial emission from discrete sources). The paper is organized as follows. In §2 we discuss spatial analysis of the ROSAT X-ray data. In §3 we discuss spatial analysis of *I*-band observations of NGC 1332. Spectral analysis of the X-ray data is presented in §4. In §5 we clarify the geometric test for dark matter and apply it to NGC 1332. Detailed analysis of the composite mass distribution and the dark matter are presented in §§6 and 7. We discuss the implications of our results in §8 and present our conclusions in §9.

## 2. Spatial Analysis of the X-ray data

NGC 1332 was observed for 25.6 ks on August 13-14, 1991 with the Position Sensitive Proportional Counter (PSPC) on board ROSAT (Trümper 1983). For a description of the ROSAT X-ray telescope see Aschenbach (1988) and for a description of the PSPC see Pfeffermann et al. (1987). Table 1 summarizes the details of the observation.

In relation to optical images of nearby early-type galaxies, X-ray images generally have much lower signal-to-noise ($S/N$). The relatively noisy X-ray data thus allows only useful aggregate constraints of the radial and azimuthal shape of the surface brightness, whereas the detailed two-dimensional surface brightness of an optical image of a typical early-type galaxy may be usefully analyzed with elliptical isophote fitting (e.g., Jedrzejewski 1987). As a result, we analyze the radial profile of the X-ray surface brightness in azimuthally averaged circular annuli (§2.2) and the ellipticity in elliptical apertures of increasing size (§2.3).

The distance to NGC 1332 has been determined by several different methods, including Hubble-flow analysis $D = 19.0$ Mpc (e.g., in Canizares, Fabbiano, & Trinchieri 1987), $D_n - \sigma$, which gives 24.1 Mpc (Donnelly, Faber, & O'Connell 1990), and surface brightness fluctuations, 20.1 Mpc (J. Tonry 1994, private communication); each distance we have scaled to $H_0 = 80 h_{80}$ km s$^{-1}$ Mpc$^{-1}$. For this paper we adopt $D = 20 h_{80}^{-1}$ Mpc for the distance to NGC 1332; at this distance $1'' \sim 0.1$ kpc.

### 2.1. Image Reduction

To prepare the X-ray image for spatial analysis we (1) excluded time intervals where the background was anomalously high, (2) corrected for exposure variations and telescopic vignetting, (3) identified and removed point sources embedded in the galactic continuum emission, and (4) subtracted the background. Steps (1) and (2) as well as the identification of point sources were performed using the standard IRAF-PROS software.

ROSAT pointed observations are partitioned into many short exposures to maximize efficiency. Unlike NGC 720 the light curve for NGC 1332 shows several spikes, all of which occur at the beginning and/or end of the individual exposures; these mostly represent scattered solar radiation. We identified the affected time intervals by extracting the emission in a $300''$ radius of the galaxy with embedded point sources masked out (see below) and then binning the observation into 100 time bins to improve $S/N$. Judging by "eye" we excluded all time bins with count rate $\geq 0.9$ counts per second. The resulting image has 21141s of accepted observing time.

We rebinned the PSPC image of NGC 1332 into $5''$ pixels corresponding to a total $1536 \times 1536$

field of pixels, which proved to optimize $S/N$ and bin-size requirements for computing the ellipticity in §2.3.1 (see BC94). Only data from the hard band (0.4 − 2.4 keV) were used in order to minimize contamination from the X-ray background and the blurring due to the point spread function (PSF) of the PSPC (see §2.2).

The flat-field correction for NGC 1332 is particularly important because the galaxy center is $7'$ off-axis, just where vignetting becomes important for the PSPC. We divided the image by the exposure map provided with the observation which corrects for both exposure variations across the field and for vignetting; note the exposure map is a factor of six coarser than our chosen pixel scale. In principle this correction depends on the energy of each individual photon, but for energies above 0.2 keV the energy dependence is small and we neglect it (Snowden et al. 1994). In Figure 1 (a) we show contours of the flat-field corrected image of NGC 1332.

The next steps to prepare the image for analysis are to identify and remove point sources embedded in the continuum emission of the galaxy. We identify sources in the field using the results from the Standard Analysis System Software (SASS) provided with the observation; SASS employs a maximum-likelihood algorithm which is explained in the *detect* package in PROS. Since, however, the software has difficulty identifying sources embedded in a continuum, we identified one source "by eye" located within $\sim 1'$ of the galaxy center in addition to the SASS sources. In all we identified five embedded sources lying in a $15' \times 15'$ box centered on the galaxy; this includes the entire region of significant galaxy emission (§2.2). The positions of the identified sources are listed in Table 2.

In order to limit the introduction of spurious features into the image shape parameters (§2.3.1), it is vital to remove the effects of the embedded sources on the galactic continuum emission. Contamination due to embedded sources is especially problematic when the continuum is nearly circular or has low $S/N$; a significant distortion will generate a preferred direction and non-zero ellipticity. We remove the embedded sources by "symmetric substitution" which is particularly suited to our analysis of the intrinsic shape of the underlying mass (see Buote & Tsai 1995a for a discussion of this technique). In essence, we replace a particular source with the continuum emission in the regions obtained by reflecting the source over the symmetry axes of the image.

We obtained the symmetry axes from the source-free region $r \leq 45''$ using the iterative moment technique described in §2.3.1. The position angle ($PA$) so obtained is 132 deg N-E and we found that the ellipticity profiles do not appreciably differ for symmetry axes within the 90% confidence limits 111 deg − 152 deg. We decided to set the orientation of the symmetry axes to the optical $PA = 115^{\deg}$ (§3) because it is much more precisely determined and it is within the uncertainties of X-ray $PA$. We display in Figure 1 (b) the central portion of the image of NGC 1332 with the sources removed.

Finally, we computed the background from examination of the azimuthally averaged radial profile; in Figure 2 we show the radial profile binned in $30''$ circular bins. Since the galactic emission only extends to $\sim 250''$ from the center a mean background level is sufficient; i.e. the cosmic X-ray background should not vary significantly over the galaxy. Computing the mean background in an annulus extending from $350'' - 400''$ centered on the galaxy centroid (§2.2) we obtained a mean background count rate of $3.22 \times 10^{-4}$ counts s$^{-1}$ arcmin$^{-2}$. Note that we will subsequently subtract the background for construction of the radial profile (§2.2) but not for computation of the ellipticity (§2.3).

## 2.2. Radial Profile

After subtracting the background, we constructed the radial profile following BC94. That is, we first computed the centroid of the X-ray emission in a circle of radius $90''$ which contains $\sim 75\%$ of the galaxy counts. An initial guess for the center was selected by "eye" and then iterated until the centroid changed by less than 0.1 pixels. We also investigated the effect of choosing different-sized radii for the circle and find that centroid position varies by $< 0.5$ pixels for radii $\leq 110''$. The centroid value is listed in Table 1. We binned the radial profile such that each circular bin has $S/N \geq 2.5$. This corresponds to $5''$ bins from $r = 0'' - 15''$, $15''$ bins from $r = 15'' - 105''$, and $45''$ bins from $r = 105'' - 285''$; we do not find an appreciable gain in $S/N$ when the circular annuli are replaced with elliptical annuli consistent with the shapes and orientations obtained in §2.3.1. The background-subtracted radial profile is displayed in Figure 3.

The PSPC off-axis PSF described by Hasinger et al. (1993; updated May 1994) depends on both the energy of the incident photon and the off-axis position. We adopted a mean value for the energy by taking a counts-weighted average of the spectrum (§4) in the hard band (0.4-2.4 keV) which yielded $\langle E \rangle = 0.82 \pm 0.36$ keV. We set the position of the PSF at the centroid of the galaxy emission which lies off-axis $\theta = 7.25'$ to the West. In Figure 3 we plot the PSF with these parameters adjusting the normalization to give a best fit to the radial profile. The PSF is too flat in the core to describe the emission of NGC 1332, even if it is a point source. The PSF is still too flat if we set $\langle E \rangle = 1.1$ keV, the energy that effectively minimizes the width of the PSF and is consistent with the PSPC spectrum (see Figure 3). By demanding that the inner $60''$ of the radial profile be fit well by the $\beta$ model (see below) we decided to use the PSF of Hasinger et al. evaluated at $\langle E \rangle = 1.1$ keV and $\theta = 5'$. We arrived at this choice as a compromise between obtaining a best fit and wanting to stay close to the real off-axis position of NGC 1332. We plot this PSF in Figure 3.

A convenient parametrization of the X-ray radial profiles of early-type galaxies is the $\beta$ model (Cavaliere & Fusco-Femiano 1976; Forman, Jones, & Tucker 1985; Trinchieri, Fabbiano, & Canizares 1986),

$$\Sigma_X(R) \propto \left[1 + \left(\frac{R}{a_X}\right)^2\right]^{-3\beta + 1/2}, \qquad (1)$$

where $a_X$ and $\beta$ are free parameters. The $\beta$ model is useful as (1) a benchmark for comparison of $\Sigma(R)$ to other galaxies and (2) an analytic parametrization for computing the gas mass (§7.2). In order to obtain physical constraints on the parameters $a_X$ and $\beta$, we convolved $\Sigma_X$ with the PSPC PSF (as described above) and performed a $\chi^2$ fit to the radial profile; note the fitted parameters $a_X$ and $\beta$ vary by less than 50% and 5% respectively over the entire range of PSFs considered above.

To make a fair comparison between model and data we evaluated the $\beta$ model convolved with the PSF on a grid of $5''$ pixels similar to the real image. We computed the radial profile binned as above and then performed a $\chi^2$ fit between the model radial profile and that of the data. The best-fit model is shown in Figure 4 and the confidence limits in Table 3.

We will find it convenient to analyze the radial profile without the three bins from $0'' - 15''$ (§6.2). For this case we rebinned the image into $15''$ pixels and computed the radial profile as before, except that the inner $15''$ is now one bin. We plot the best-fit model (as above) in Figure 4 and give the confidence limits of the parameters in Table 3.

### 2.3. Ellipticity of the X-ray Surface Brightness

#### 2.3.1. Methods and Results

Like NGC 720, NGC 1332 has many fewer counts ($\sim 1000$ for $r \lesssim 100''$) than typical optical images of nearby ellipticals. As a result, we can only hope to measure with any precision the aggregate elongation of the X-ray surface brightness in a large aperture. The iterative moment technique introduced by Carter & Metcalfe (1980) is particularly suited to measuring aggregate shapes; see BC94 for specific application to X-ray images. In essence this technique entails computing the analog of the two-dimensional moments of inertia arrived at by iterating an initially circular region; the square root of the ratio of the principal moments is the axial ratio and the orientation of the principal moments yields the position angle. The parameters obtained from this method, $\epsilon_M$ and $\theta_M$, are good estimates of the ellipticity ($\epsilon$) and position angle ($\theta$) of an intrinsic elliptical distribution of constant shape and orientation. For a more complex intrinsic shape distribution, $\epsilon_M$ and $\theta_M$ are average values weighted heavily by the outer parts of the region.

BC94 found that the uncertainties on $\epsilon_M$ and $\theta_M$ obtained from Monte Carlo experiments were characterized well by their analytical estimates, $\Delta\epsilon_M$ and $\Delta\theta_M$, computed from simple propagation of Poissonian errors (see appendix A of BC94). For NGC 720, however, the X-ray emission is more extended and the $S/N$ is much larger than for the NGC 1332 PSPC data. We find, thus, that the analytic statistical error estimates generally underestimate the uncertainties for NGC 1332. As a result, we estimated the uncertainties by simulating $\beta$ models having the best-fit $a_X$ and $\beta$ obtained in §2.2 but also have a constant ellipticity and orientation; we implemented this by replacing $R = \sqrt{x^2 + y^2}$ with the elliptical radius $\sqrt{x^2 + y^2/q^2}$, where $q$ is the constant axial ratio. These models were scaled to have the same number of counts as the background-subtracted

PSPC image of NGC 1332; then a uniform background (scaled to the PSPC observation) and Poisson noise was added. Since $\epsilon_M$ is unaffected by a uniform background (see Carter & Metcalfe 1980) we did not then subtract an estimate for the background. We performed 1000 simulations each for a suite of ellipticities; note that we do not include the PSF in these simulations – the PSF is taken into account in our models in later sections.

To determine the confidence intervals on the measured $\epsilon_M$ we proceed as follows; for now we focus our attention on the 90% confidence limit for a particular aperture size. We arrange the results of the 1000 simulations for a given input $\epsilon_x$ into ascending order of measured ellipticity $\epsilon_M$; i.e. $\epsilon_M^1 < \epsilon_M^2 < \cdots < \epsilon_M^{1000}$. The 90% upper limit for this model is defined to be the value of $\epsilon_M$ corresponding to the $0.9 \times 1000 = 900$th value of $\epsilon_M$ in the ordered array; i.e. $\epsilon_M^{900}$. The 90% confidence lower limit of the real data is given by the model with input $\epsilon_x$ whose $\epsilon_M^{900}$ just equals the measured value of $\epsilon_M$ from the real image (see below). Similarly, the 90% confidence upper limit of the real data is given by the model with input $\epsilon_x$ whose $\epsilon_M^{100}$ just equals the measured value of $\epsilon_M$ from the real image. The same procedure applies to different confidence limits and aperture sizes.

We estimated the confidence limits for the position angles $\theta_M$ from the model having $\epsilon_x = \epsilon_M$ of the data for the specific aperture size. We define, for example, the 90% confidence limits to be $(\theta_M^{50}, \theta_M^{950})$; the other confidence limits are defined similarly. This procedure is not the most rigorous means to determine the uncertainties on $\theta_M$ because the value of $\epsilon_M$ in each simulation is not equal to $\epsilon_x$. Since, however, we do not use $\theta_M$ in our modeling (§6) the estimates are sufficient for our purposes.

In addition to the iterative moments, we also parametrized the shape of the X-ray surface brightness by fitting perfect ellipses to the isophotes following Jedrzejewski (1987; implemented with the *ellipse* task in the IRAF-STSDAS software). This method has the advantage that the computed parameters for ellipticity ($\epsilon_{iso}$) and position angle ($\theta_{iso}$) correspond to an elliptical isophote at a specific radius and thus may provide a more accurate representation of the radial variation in shape and orientation of the surface brightness. Unfortunately this technique, which was developed to study slight departures of optical isophotes from true ellipses, has the disadvantage of having larger statistical uncertainties than the iterative moments; i.e. as applied in STSDAS, the counts associated with fitting an isophote are only a small fraction of those present in the elliptical apertures used to compute the iterative moments. To improve the $S/N$ we rebinned the image into $15''$ pixels for the ellipse fitting.

We list the results for $\epsilon_M$ and $\epsilon_{iso}$ in Table 4 and the results for $\theta_M$ and $\theta_{iso}$ in Table 5; note that the ellipticities in Table 4 include the blurring due to the PSPC PSF (which we will account for in our models in the later sections) and the background, since (as mentioned above) $\epsilon_M$ is unaffected by a uniform background; also, only 68% confidence limits are given for the results from isophote fitting. The ellipticities for the two methods agree within their considerable uncertainties at all radii. As expected, the iterative moment results are determined more precisely. At the 90% confidence level, $\epsilon_M$ is consistent at all radii, although at 68% confidence the innermost $a \leq 30''$ and outermost $r \geq 135''$ are rounder than the intermediate radii; this simply reflects the blurring due to the PSF for the inner isophotes and the low $S/N$ in the outer isophotes which do not allow the iterative moment method to iterate much past its initial guess of a circle.

The most accurately determined values for $\epsilon_M$ are for $a \sim 75'' - 90''$ giving $\epsilon_M = 0.10 - 0.27$ at 90% confidence. These results are similar to those for NGC 720 obtained by BC94 where at the aperture sizes $a = 90''$, $\epsilon_M = 0.20 - 0.30$ at 90% confidence. Unlike NGC 720, there is no evidence for significant elongation at large radii (i.e. $a \geq 90''$) because of the sharp falloff in $S/N$ at those radii in NGC 1332.

There is no evidence for position angle twists at the 90% confidence level. As a result we define the X-ray position angle, $\theta_X$, at the radii where $\epsilon_M$ and $\theta_M$ are best-determined; i.e. $a \sim 75'' - 90''$. The mean value of $\theta_M$ agrees well with the optical position angle (see §3) of 115 deg. Thus we define $\theta_X \equiv \theta_{opt} = 115$ deg.

### 2.3.2. Search for Unresolved Sources

Following BC94 we investigated the possibility that the measured values of ellipticities and position angles are due to unresolved point sources embedded in the galactic emission; note Buote & Tsai (1995b) use simulations to show that the effects of unresolved point sources in ROSAT PSPC images of galaxy clusters on measurements of surface brightness shape parameters are generally comparable to or less than the Poisson noise. To search for unresolved sources, we examined (1) the centroid variation with radius, (2) the symmetry of surface brightness, (3) subtraction of a model for the galaxy continuum, and (4) one-dimensional projections along the major and minor axes.

Methods (1), (3), and (4) yield null results. The centroids for both the iterative moments and the isophote fitting varied by less than 0.5 pixels (i.e. $2.5''$) for $a \leq 150''$. For (3) we used the results of the isophote fits to construct a model for the surface brightness within $r = 100''$ (see *bmodel* task in IRAF-STSDAS) and then subtracted the model from the image. The residuals showed no statistically significant fluctuations except at the center $r < 30''$ where the model is a poor fit to the data (i.e. model is too flat). For (4), we examined one-dimensional projections of the X-rays along the major and minor axes in boxes of width $150''$, $200''$, and $250''$. In each case there were no statistically significant asymmetries.

Method (2) yielded possible evidence for asymmetries. We considered the surface brightness in a coordinate system where the $x$-axis was aligned with $\theta_X$. We constructed an image from the first quadrant by reflecting it over the $x$ and $y$ axes; we did the same for the other quadrants to give four images, one derived from each quadrant. If the X-rays possess elliptical symmetry each of these images should possess the same shape. On computing $\epsilon_M$ from these images we found that the values were quite different for the different quadrants. For apertures $a = 75'' - 90''$, we obtained $\epsilon_M \sim 0.15$ for the first and fourth quadrants and $\epsilon_M \sim 0.30$ for the second and third quadrants. Judging from the Monte Carlo error estimates in §2.3.1 these values are certainly consistent within their 90% confidence values. Of course, it is difficult to precisely ascertain the significance of these values because each quadrant only has a fraction of the counts we have now given it for the whole image. We conclude that asymmetries are consistent with the data, but are not demanded statistically by the data.

## 3. Spatial Analysis of the Optical data

In later sections we require a mass model for NGC 1332 obtained from assuming a constant mass-to-light ratio. Hence, on August 25, 1992 we obtained a 250s $I$-band exposure of NGC 1332 with the 1.3m McGraw-Hill telescope at the Michigan-Dartmouth-MIT observatory (MDM). The image was recorded on the Thomson CCD which has a $400 \times 576$ field of $0.5''$ pixels. From stars in the field we estimated the seeing to be $2.0''$ FWHM.

We reduced the image using the IRAF-NOAO software. First, we subtracted a constant bias level computed from un-illuminated columns at the edge of the chip. The image was flattened using the average of a suite of sky flats (see Haimen et al. 1994). Next, we subtracted a constant sky value. Finally, we fitted elliptical isophotes as described in §2.3.1.

In addition to this short-exposure observation which emphasizes the core of the galaxy, we also obtained two-dimensional isophotal photometry for a deep exposure of NGC 1332 (J. Tonry 1995, private communication). This 900s $I$-band exposure was taken on October 10, 1991 with the 4m reflector at CTIO in fair seeing conditions ($1.41''$ FWHM). Although this image is especially suited to analysis of the outer regions of the galaxy, the core ($r \lesssim 5''$) is unfortunately saturated.

In Table 6 we list the ellipticities and position angles for the two data sets; note that two-dimensional isophotal surface photometry has not been published in the literature for NGC 1332. The data sets agree quite well within the uncertainties computed for the MDM data; we do not have uncertainties for the Tonry observation. The isophotes are round near the center $a \lesssim 5''$ where the PSF dominates and become increasingly elongated with distance from the center eventually reaching $\epsilon_{opt} \sim 0.70$ at $a \sim 75''$. The position angles are quite steady at $\theta_{opt} \sim 115 \deg$ over most of the radii observed, but for $a < 2''$ for the MDM data and $a > 130''$ for the Tonry data there may be significant twists. The outer twist in the Tonry data may be due to its dwarf companion NGC 1331; the inner twist is not quantifiable due to the seeing.

We combined the major-axis intensity profiles from the two data sets by minimizing the deviations in their overlapping data points; the ellipticity profiles were taken to be those of the MDM data for $a \leq 75''$ and the Tonry data for $a > 75''$. In Figure 5 we plot the combined major-axis intensity profile for the two data sets. The intensity-weighted average ellipticity over the listed radii is $\langle \epsilon_I \rangle = 0.43$, which is necessarily affected to some extent by the blurring of the inner isophotes by the PSF.

The intensity profile exhibits two breaks: one near $a = 1''$ resulting from the MDM PSF and one near $a = 40''$ where apparently the light profile of the disk becomes important. We parametrized the major-axis profile using a simple bulge + disk model consisting of a De Vaucouleurs bulge and an exponential disk,

$$\Sigma_I(R) \propto \exp\left(-7.67\left[(R/R_e)^{1/4} - 0.5\right]\right) + C\exp\left(\frac{-R}{R_d}\right), \qquad (2)$$

where the best-fit parameters are the effective radius, $R_e = 25''$, the disk scale length, $R_d = 33''$, and a relative normalization parameter $C = 2.6$; these parameters were obtained by fitting the bulge + disk model convolved with the PSF of the MDM observation taken to be a Gaussian ($\sigma = 0.85''$). We plot this model in Figure 5; also plotted are the results of a single De Vaucouleurs Law fit to the whole profile ($R_e = 55''$) that will prove more convenient for analytic modeling.

We investigated whether the optical surface brightness profile is consistent with the X-ray emission by fitting the bulge + disk model to the X-ray radial profile (§2.2). We constructed a two-dimensional optical model by replacing $R = \sqrt{x^2 + y^2}$ with the elliptical radius $\sqrt{x^2 + y^2/q^2}$ in equation (2), where $q = q(R)$ is the axial ratio of the optical isophote at major-axis position $x = R$; note we interpolate between the discrete values listed in Table 6. Following our procedure for the $\beta$ model in §2.2, we convolved the elliptical bulge + disk model with the PSPC PSF and evaluated it on a grid of $5''$ pixels. Then the radial profile is computed in the same manner as for the X-ray data and is compared to the X-ray radial profile. We only needed to rescale the optical profile since all of the parameters are fixed to the optical values. The result is shown in Figure 5 – $\chi^2 = 190$ for 11 degrees of freedom. The poor fit clearly demonstrates that a large fraction of the X-ray emission must not be distributed like the optical light.

We also used the above model evaluated on a grid of $5''$ pixels and convolved with the PSPC PSF to give the range of $\epsilon_M$ expected if the X-ray gas follows the optical light. Following the procedure of §2.3.1 we computed $\epsilon_M$ for 1000 Monte Carlo simulations. For $a \leq 90''$, the 90% lower limits for $\epsilon_M$ (i.e. $\epsilon_M^{100}$) generated by the optical model generally agree with the X-ray values in Table 4. However, at larger radii the optical model predicts $\epsilon_M$ larger than observed at $> 95\%$ confidence.

## 4. Spectral Analysis of the X-ray data

The ROSAT PSPC has 256 PI bins spanning the energy range 0.1 - 2.4 keV. Only four groups of bins, however, are resolved in energy. Since NGC 1332 was observed in August, 1991, we used the response matrix and off-axis parameters appropriate to AO1 observations; we always excluded PI bins 1-10 and 224-256 from analysis due to calibration uncertainties (ROSAT Status Report # 78). We extracted the source spectrum using the IRAF-PROS software.

First, we analyzed the reduced image (§2) to determine the optimal region to extract the source. After experimenting with circles of different radii centered on the galaxy centroid (§2.2), we selected $R = 120''$ as that radius which optimized the $S/N$ of the background-subtracted emission. We extracted the spectrum in this region using the full-scale PSPC image ($15360 \times 15360$ field of $0.5''$ pixels) corrected only for bad time intervals and embedded point sources; note the sources were simply masked out, not symmetrically substituted (§2). The background level was computed in an annulus $R = 350'' - 400''$ and then subtracted from the extracted spectrum after correcting for telescopic vignetting. To improve $S/N$ we rebinned (using *grpha* in FTOOLS) the resulting spectrum into 15 spectral bins each having $> 20$ counts. With XSPEC, we fit the background-subtracted spectrum to a single-temperature ($1T$) optically thin plasma incorporating thermal bremsstrahlung and line emission (Raymond & Smith 1977; updated to 1992 version) with Galactic absorption. The temperature, metallicity, hydrogen column density, and emission normalization were free parameters in the fits.

Table 7 summarizes the spectral data and fit results. The Raymond-Smith $1T$ model fits the data quite well. The 68% and 90% confidence levels for the three interesting parameters (temperature, abundances, and $N_H$) are determined by the contours of constant $\chi^2_{min}+$ 3.53 and 6.25 respectively; these contours are plotted in Figure 6. The constraints on $N_H = (0.083 - 4.7) \times 10^{20}$ cm$^{-2}$ (90% confidence) are consistent with the galactic column density $N_H = 2.2 \times 10^{20}$ cm$^{-2}$ (Stark et al. 1992). The abundances, where He was fixed at its cosmic value and the heavy element abundances have relative abundances fixed at solar, are not well determined, but the temperature is constrained to about a factor of 2. Like BC94 find for NGC 720, we find that two-temperature models having abundances fixed at solar fit the spectrum equally well.

We investigated the presence of temperature gradients by employing the same fitting procedure and single-temperature models as above. For examination of radial gradients we partitioned the $120''$ region into an inner circle ($30''$) and an outer annulus ($30'' - 120''$) each possessing roughly the same $S/N$; although the $r = 30''$ region is only slightly larger than the size of the PSF and thus undoubtedly has some correlation with the outer region, the constraints on the temperature gradient are hardly affected by selecting a larger inner circle — the outer region becomes more unconstrained because of the corresponding decrease in size and thus $S/N$. In any event, when comparing temperature gradients produced from the hydrostatic models (§6.2.2) to the PSPC data *we always fold the PSPC PSF into the model temperature map*. We also needed to rebin the spectrum into only 8 channels to obtain acceptable $S/N$ in each bin. The spectra are plotted in Figure 7 and the results of the fit are listed in Table 7 with only 68% confidence estimates because of the greater uncertainty due to the smaller number of counts in each region. From consideration of the 68% confidence extremes, we constrain the gradient to be $-0.55 < \left(\frac{d\ln T_{gas}}{d\ln R}\right) < 0.64$ ($\sim 95\%$ confidence), where we have taken intensity-weighted values of $R$ for each of the regions. If we fix $N_H$ to its Galactic value, we obtain $-0.45 < \left(\frac{d\ln T_{gas}}{d\ln R}\right) < 0.49$ at $\sim 95\%$ confidence.

In order to set more stringent limits on radial temperature gradients, we applied a Kolmogorov-Smirnov (K-S) test to the spectra of the two regions. Since the K-S test is intended for analysis of un-binned data, we used the full-scale spectrum consisting of 256 PI bins (with bins 1-10 and 224-256 excluded as above). In addition, since our spectra contained fractional counts (due to background subtraction) we rounded off fractions to the nearest whole number; this rounding off was especially important because after the background was subtracted the many bins that had had only one count then had slightly less $\sim 0.9$ which would be unfairly eliminated by simply truncating the fractions.

The results of the K-S tests depended on the energy range examined. When channels $\lesssim 0.2$ keV were included, we obtained probabilities $P_{KS} \ll 1\%$ that the two regions are derived from the same population. However, the X-ray background contribution is largest in the low-energy bins and, because it covers a larger area, the outer region certainly suffers more contamination from errors in the background determination than the inner region; i.e. as a result of imperfect background subtraction the background in each region is a different proportion of the total spectrum which necessarily affects the shape of the cumulative probability distribution responsible for $P_{KS}$. To reduce background contamination we restricted analysis to the hard band (0.4 - 2.0 keV). We found then that $P_{KS} = 40\%$ for energies 0.4 - 2.0 keV and that $P_{KS}$ varied by $\lesssim 10\%$ for other nearby cuts of energy; e.g., $P_{KS} = 34\%$ (0.3 - 2.0 keV) and $P_{KS} = 29\%$ (0.5 - 2.0 keV).

As we found for NGC 720 (BC94), this large probability serves as a discriminator for models with steep temperature gradients under the assumption that differences in the spectra of the two regions are due primarily to differences in temperature. To see how sensitive such a test would be for detecting a real temperature gradient we simulated Raymond-Smith spectra (with XSPEC) with statistics appropriate for the PSPC observation of NGC 1332; in each region (i.e. $R = 0'' - 30''$ and $R = 30'' - 120''$) the simulated spectra have Galactic $N_H$ and 50% metallicities but different temperatures. We found that for a temperature in the inner region of $T_{in} = 0.60$ keV and an outer region temperature of $T_{out} = 0.50$ keV, $P_{KS} = 37\%$. The probability fell to $P_{KS} = 15\%$ for $T_{in} = 0.60$ and $T_{out} = 0.45$. However, for a slightly larger gradient (i.e. $T_{in} = 0.65$ keV, $T_{out} = 0.40$ keV), the probability dipped below 1%; this held for other $T_{in}$ and $T_{out}$ between $0.4 - 1.0$ keV. We defined spectral models to be inconsistent with the data if $P_{KS} < 1\%$. With this criterion, we determined that for reasonable temperature ranges (i.e. $0.4 - 1$ keV), the temperature gradient is constrained to be $\left| \frac{d \ln T_{gas}}{d \ln R} \right| < 0.35$. Hence, we find no evidence for radial temperature variations, but the PSPC spectrum cannot rule out sizeable gradients.

Following BC94 we also tested for azimuthal gradients by slicing the $120''$ circle into 4 equal wedges of $90°$. We defined the edges of the wedges with respect to the major axis to be (1) $-45 \deg$ to $+45 \deg$, (2) $+45°$ to $+135 \deg$, (3) $+135 \deg$ to $+225 \deg$, and (4) $+225 \deg$ to $-45 \deg$; the major axis was taken along P.A. 115 deg. We grouped wedges (2) and (4) into a region denoted (A) and regions (1) and (3) were grouped into region (B) in order to improve the $S/N$. The results of the fits for these regions are listed in Table 7. A K-S test of (A) and (B) (0.1 - 0.3 keV bins omitted) yields a probability of 50% that the two regions are derived from the same population. Hence we find no evidence for a temperature gradient between (A) and (B).

Finally, we considered a possible contribution to the emission from discrete sources in the galaxy. We modeled the spectrum of discrete emission by (1) a power law, (2) a $T = 8$ keV Bremsstrahlung component, and (3) a $T = 8$ keV Raymond-Smith plasma with Galactic $N_H$ and 50% metallicities. None of these spectra alone fit the PSPC spectrum very well and can be ruled out as the dominant component of the X-ray emission (see Table 7); this is consistent with the poor fit of the optical radial profile to the X-ray radial profile in §3. However, the PSPC data cannot precisely constrain the relative flux of the hot gas and discrete components; i.e. equal contributions of hot gas and discrete emission are allowed by the spectrum. We discuss the implications of emission from discrete sources in §§5.2.2, 6.2.2, and 6.3.

## 5. Geometric Test for Dark Matter: Clarification and Applications

### 5.1. Theory

Here we clarify the test for dark matter in early-type galaxies introduced by BC94 that employs only the observed optical and X-ray surface brightness distributions of the galaxies and is completely independent of the temperature profile of the gas. We now show that this geometric test in principle allows for a model-independent test for whether mass traces light in a galaxy. More generally, any mass distribution may be tested for consistency with the X-ray data. First, the arguments of BC94 are summarized.

The fundamental assumption of the geometric test is that the X-ray emission may be approximated as due to hot gas in quasi-hydrostatic equilibrium with the underlying gravitational potential of the galaxy,

$$\nabla p_{gas} = -\rho_{gas} \nabla \Phi, \tag{3}$$

where $\rho_{gas}$ is the gas density, $\Phi$ is the gravitational potential of the galaxy, and $p_{gas}$ is the gas pressure. By "quasi" we mean that any streaming and rotation present must be unimportant with respect to the thermal pressure and gravitational potential energy of the galaxy. It is simple, however, to incorporate any measured rotation of the gas by replacing $\Phi$ with the appropriate effective potential $\Phi_{eff}$.

The hydrostatic equation generically requires that $\Phi$, $\rho_{gas}$, and $p_{gas}$ are all stratified on the same constant surfaces in three dimensions. With the additional assumption of a single-phase gas, the temperature, $T_{gas}$, must also share the same constant surfaces. In order to prove this property, we take the the curl of equation (3) and obtain $\nabla \rho_{gas} \times \nabla \Phi = 0$; i.e. $\rho_{gas}$ and $\Phi$ have parallel normal vectors which means that $\rho_{gas}$ and $\Phi$ are stratified on the same constant surfaces in three dimensions. If we instead divide equation (3) by $\rho_{gas}$ and take the curl we obtain (after some rearranging) that $\nabla \rho_{gas} \times \nabla p_{gas} = 0$; i.e. $\rho_{gas}$ and $p_{gas}$ share the same constant surfaces in three dimensions. If $p_{gas} = p_{gas}(\rho_{gas}, T_{gas})$ it follows that on surfaces where $\rho_{gas}$ is constant, $T_{gas}$ must also be constant. Hence the hydrostatic equation alone demands that $\Phi$, $\rho_{gas}$, $p_{gas}$, and $T_{gas}$ all share the same constant surfaces in three dimensions. Note that no assumption about the form of the pressure or temperature is required for this property except that the gas is adequately described by a single phase; also note that an ideal gas need not be assumed as in BC94.

The important observable quantity, the gas volume emissivity, is also stratified on the same constant surfaces because it is a function only of the gas density and temperature: $j_{gas} \propto \rho_{gas}^2 \Lambda_{PSPC}(T_{gas})$, where $\Lambda_{PSPC}$ is the plasma emissivity convolved with the spectral response of the PSPC; $\Lambda_{PSPC}$ is only a weak function of $T_{gas}$ and the metallicity of the gas (e.g., the poor constraints on these parameters obtained for NGC 4636 by Trinchieri et al. 1994). Hence $j_{gas}$ and $\Phi$ trace exactly the same shape in three dimensions, regardless of the temperature profile of the gas; note that if the gas is rotating, $j_{gas}$ traces the same shape as $\Phi_{eff}$. This correspondence between the shapes of $j_{gas}$ and $\Phi$ is the basis for the geometric test for dark matter.

*We now depart from the presentation of §3.1 of BC94* who discussed the qualitative similarity between the shapes of the contours of the projected potential and the X-ray isophotes. Instead we clarify and extend the procedure employed in §3.2 of BC94 to test whether the optical light traces the underlying mass. In BC94 we implied that such a comparison was qualitative (because of projection effects), now *we show that the comparison is rigorous.*

To ascertain whether light traces mass we first obtain $\Phi$ from the optical surface brightness distribution $\Sigma_L$. We make the assumption that the three-dimensional luminosity density, $j_L$, is axisymmetric with its symmetry axis inclined by angle $i$ with respect to the line of sight. With the additional general assumption that $j_L$ may be represented as a finite series of spherical harmonics, Palmer (1994) has shown that $j_L$ may be uniquely and analytically deprojected from $\Sigma_L$. For mass tracing light the mass density is just $\rho_L \propto j_L$ and the corresponding potential $\Phi_L$ follows from Poisson's equation. Setting $\Phi = \Phi_L$ we have the potential under the assumption mass traces light. By applying the same assumptions for $j_L$ to $j_{gas}$, then $j_{gas}$ may be uniquely and analytically deprojected from the X-ray surface brightness $\Sigma_X$. If $\Phi$ and $j_{gas}$ trace the same shape, then mass tracing light is a suitable description of the galaxy. *Inconsistency of the shapes of $\Phi$ and $j_{gas}$ signals the presence of dark matter, independent of the temperature profile of the gas and without requiring any model fitting.*

Apart from the fundamental assumption of hydrostatic equilibrium, the few additional assumptions of this geometrical test for dark matter are not restrictive. First, by examining both oblate and prolate axisymmetric deprojections, one necessarily brackets the full range of ellipticities of triaxial galaxies. Second, the unknown inclination angle must be the same for $\Phi$ and $j_{gas}$ if mass traces light. Since an inclined axisymmetric galaxy is necessarily rounder in projection than when viewed edge-on, deprojection assuming $i = 90$ deg will yield $\Phi$ and $j_{gas}$ rounder than the true values. This serves only to make any deviations in the shapes of $\Phi$ and $j_{gas}$ more difficult to observe. Third, the assumption that $j_{gas}$ and $j_L$ be represented as a finite series of spherical harmonics should bracket most physical cases of interest. Finally, the assumption of a single, dominant gas phase should be a good description except in the regions of a strong cooling flow (Thomas, Fabian, & Nulsen 1987; Tsai 1994).

The standard X-ray method for analyzing the mass distribution in early-type galaxies utilizes the spherically-symmetric solution of equation (3) for a single-phase ideal gas (for a review see Fabbiano 1989). This method emphasizes the radial mass distribution and suppresses any information contained in the shapes of the X-ray isophotes. The mass within a radius $r$ for this solution is expressed in terms of $d\log\rho_{gas}/d\log r$, $T_{gas}(r)$, and $d\log T_{gas}/d\log r$. As a result, employing this solution requires detailed information on the intrinsic temperature profile of the gas. Moreover, since $\rho_{gas}$ appears explicitly it must be disentangled from the observed surface brightness ($\Sigma_X$) assuming knowledge of the plasma emissivity $\Lambda_{PSPC}$ of the gas. In contrast, the geometric test proposed by BC94, which emphasizes the elongation of the mass, requires only $j_{gas}$ and hence does not depend on knowledge of the temperature profile or the plasma emissivity of the gas. (Note that this test does *not* give any information regarding the radial mass distribution, although it can give the ratio of dark matter to luminous matter.)

BC94 performed the geometric test for dark matter on NGC 720 similar in concept, but different in implementation, to what we have described above. The spatial resolution of the PSPC ($FWHM \sim 30''$) does not warrant detailed deprojection of $\Sigma_X$ on scales much smaller than $30''$. Since the isophotes of interest for BC94 lie at $\sim 100''$ and the PSPC image of NGC 720 has relatively low $S/N$ ($\sim 1500$ counts), we opted for a coarser comparison of $\Phi$ and $j_{gas}$. BC94 deprojected $\Sigma_L$ to obtain $j_L$ by fitting a simple model to the major-axis of $\Sigma_L$ and assigning to $j_L$ the ellipticity of the flattest isophotes of $\Sigma_L$. By assuming mass follows light BC94 computed $\Phi_L$ from $j_L$. Invoking hydrostatic equilibrium, they assigned the ellipticity profile of $\Phi_L$ to $j_{gas}$. They then used a simple parametrization of $j_{gas}$ as a spheroid having ellipticity varying with radius to generate the corresponding $\Sigma_X$ consistent with the PSPC radial profile. Finally, the ellipticity of this $\Sigma_X$ consistent with $\Phi_L$ was compared to the ellipticity of the data. By performing 1000 Monte Carlo realizations of $\Sigma_X$ BC94 concluded that the X-ray isophotes generated assuming mass traces light are rounder than the observed isophotes at the 99% confidence level. Note that BC94 implied that the comparison was uncertain due to projection effects whereas we have shown here that the comparison is in fact more robust.

### 5.2. Application

Because the important apertures ($a \sim 80''$) for comparison of $\epsilon_M$ lie nearer to the center for NGC 1332 than for NGC 720 ($a \sim 100''$) and the PSPC image of NGC 1332 has even lower $S/N$ than that of NGC 720, we apply the geometric test using simple parametrizations of $\Sigma_L$ and $\Sigma_X$ following BC94. Moreover we address the effects of possible rotation of the gas and contamination of the X-ray emission due to discrete sources in the galaxy. (In BC94 we did not need to explore in detail these effects for NGC 720 because (1) there is negligible measured rotation in the optical for NGC 720 (Binney, Davies, & Illingworth 1990), and (2) the PA offset of the X-ray and optical isophotes argues against a significant contribution to the X-ray emission from stellar (discrete) sources.) We assume that a single-phase gas adequately describes the bulk emission.

Before embarking on the robust geometrical test, we obtain a useful qualitative picture by comparing the shapes of the X-ray isophotes (§2.3.1) to the contours of constant projected potential for a constant $M/L$ model. BC94 showed that for a wide class of physical models the ellipticities of the projected potential and X-ray isophotes should agree to within $\Delta \epsilon \lesssim 0.04$. We obtain $\rho_L$ by deprojecting the single De Vaucouleurs profile obtained from §3 approximated by a Hernquist profile (see §6.2.1). Applying Poisson's equation we obtain the potential $\Phi_L$. Listed in Table 8 are the ellipticities of the isopotential surfaces of $\Phi_L$, its projection onto the sky assuming oblate symmetry, and its projection convolved with the PSPC PSF. We also show in Table 8 the ellipticities of the X-ray isophotes predicted by explicit solution of the hydrostatic equation for an isothermal ideal gas (see §7.3).

For semi-major axes $75'' - 90''$, those being the apertures for which $\epsilon_M$ is best determined, the projected potential convolved with the PSPC PSF and the isothermal model have ellipticities about 0.10. These values are inconsistent with the measured lower limits for $\epsilon_M$ at 68% confidence but marginally consistent at 90% confidence. For prolate constant $M/L$ models the ellipticities are virtually identical because the distinction between prolate and oblate spheroids for such small ellipticities is negligible. Considering the uncertainty $\Delta \epsilon \lesssim 0.04$ for comparing the projected potential and X-ray isophotes (BC94) the stellar mass model appears inconsistent at the 68% confidence level on $\epsilon_M$, but consistent at 90% confidence.

We make this comparison robust following the procedure of BC94 discussed at the end of §5.1. Our model for $\Phi_L$ is the same as previously discussed. We follow BC94 and parametrize $j_{gas}$ by the pseudo-spheroids discussed in Appendix B of BC94. These spheroids are a function of,

$$\zeta^2 = x^2 + y^2 + z^2/q(r)^2, \tag{4}$$

where $q = q(r) = 1 - \epsilon(r)$ is the radially varying axial ratio with $r^2 = x^2 + y^2 + z^2$; such models are very similar to those introduced by Ryden (1990). We parametrize the emissivity by a power law, $j_{gas} \propto (R_c^2 + \zeta^2)^{-n}$, with core radius $R_c$ and slope parameter $n$. The surface brightness generated by this parametrization, $\Sigma_X^{param}$, is computed by integrating $j_{gas}$ along the line of sight and then convolving with the PSPC PSF. We determine $R_c$ and $n$ by fitting the radial profile of $\Sigma_X^{param}$ to $\Sigma_X$ as discussed in §2.2; the best-fit parameters are $R_c = 0.6''$ and $n = 1.38$. Then Poisson counts and a uniform background are added to the image appropriate for the NGC 1332 PSPC observation; we do not bother to then subtract a background estimate because it does not affect computation of $\epsilon_M$ (see §2.3.1).

In Table 9 we list the upper limits on $\epsilon_M$ for 1000 simulations for both oblate and prolate deprojections; e.g., 90% upper limit is $\epsilon_M^{900}$ (see §2.3.1). For both cases, just like the qualitative comparison made above, the assumption of mass tracing light yields $\epsilon_M$ values inconsistent with the data at the 68% confidence level and marginally consistent at the 90% level. In Figure 8 we plot the mean $\epsilon_M$ from the oblate simulations superposed on $\epsilon_M$ measured from the data. Clearly it is imperative to obtain higher $S/N$ data to determine whether there is a real discrepancy.

### 5.2.1. Rotation

In the previous discussion we have neglected possible rotation of the gas. The PSPC, of course, does not have the spectral resolution or sensitivity to detect rotation so we must resort to indirect arguments to estimate its effects (see BC94 for a discussion). Since the gas mass of NGC 1332 is only a small fraction of the stellar mass (see §7.2) it is consistent with being produced by normal stellar mass loss over $\sim 10^9$ years (Mathews 1990). In such a scenario the gas should have originally rotated in a manner similar to the stars. After being heated by supernovae the gas would expand, maintaining the same angular momentum of the stars, but with a reduced velocity. In the central regions of a cooling flow, however, the velocity could be driven up above the stellar rotation rate as the gas falls back in to the center (see, e.g., Kley & Mathews 1995). The effect of such rapidly rotating gas resulting from infall cannot be substantial at the relatively large radii $r = 75'' - 90''$ that are of interest to us since only a small fraction of the stellar mass exists at larger radii. We examine the case where the gas rotates like the stars which should serve as a conservative upper limit in light of the above scenario.

Incorporating rotation into our formalism simply entails replacing the gravitational potential in eq. [3] with an effective potential $\Phi_{eff} = \Phi - \Phi_{rot}$, where $\Phi_{rot}$ is the potential due to rotation. Dressler & Sandage (1983) measured the major-axis rotation curve for NGC 1332 out to $R = 60''$. The velocity profile rises to $R = 20''$ then flattens out to a maximum velocity of $V_{max} = 230$ km s$^{-1}$. A simple means to incorporate the effect of the gas rotating like the stars is to assume the mean azimuthal rotation, $\bar{v}_\phi$, is a fraction of the circular velocity scaled to the observed $V_{max}$ as follows. For a spherical Hernquist model the circular velocity is $v_c = \sqrt{GMr}/(r + r_c)$, where $r$ is the radius and $r_c$ the core parameter. We take the velocity profile along the major-axis to have this form which becomes $\bar{v}_\phi(R,0) = 2V_{max}\sqrt{R_c R}/(R_c + R)$, where $V_{max} = \bar{v}_\phi(R_c, 0)$. Replacing $R$ by the spheroidal radius $\sqrt{R^2 + z^2/q^2}$ (with $q$, the constant axial ratio of the mass, taken to be 0.57 – see §3) this velocity profile approximately corresponds to $\Phi_{rot} = 4R_c V_{max}^2/(R_c + \sqrt{R^2 + z^2/q^2})$. Although this relation is not exact, it provides a convenient exploration of the effects of the gas rotating like the stars. To properly normalize $\Phi_{rot}$ with respect to $\Phi$ we need to specify the total mass. From the results of §6.3 we take the mass to be $0.5 \times 10^{12} M_\odot$. For ranges of plausible masses $(0.2 - 0.8) \times 10^{12} M_\odot$ appropriate to the Hernquist model the precise choice of mass is not important.

We list in Table 10 the 68% and 90% upper limits on 1000 simulations of the oblate mass models having the type of rotation just described; we consider only oblate models since minor-axis rotation is rare. Again we focus on semi-major axes $a = 75'' - 90''$ since they are best determined by the data. Even with rotation the 68% confidence upper limits fall below the observed values, although the 90% upper limits are consistent with the data; we note that at the 85% level the $a = 75'' - 90''$ upper limits are just marginally consistent with the data. *Thus rotation of the gas similar in nature to the stars does weaken the need for dark matter based solely on shape alone, but the ellipticities are only about 10% larger than the non-rotating case.* This should serve as a conservative upper limit of the effects of rotation if the gas has expanded (due to heating) and conserved angular momentum.

### 5.2.2. Emission from Discrete Sources

Finally we consider the effects of emission from discrete sources in the galaxy. We model the X-ray surface brightness ($\Sigma_{ds}$) due to discrete sources by assuming that $\Sigma_{ds} \propto \Sigma_L$, where $\Sigma_L$ is the optical surface brightness convolved with the PSPC PSF (see §3). Thus in this case our composite model surface brightness is $\Sigma_X = \Sigma_{hg} + \Sigma_{ds}$, where $\Sigma_{hg}$ is the surface brightness of the hot gas computed in §5.2. The relative normalization of the two components is given by $F_{ds}/F_{hg}$ – the ratio of the X-ray fluxes of the two components within $R = 300''$.

In Table 10 we display the results of the simulations for $F_{ds}/F_{hg} = 1/10$ and $1/3$; we consider only oblate models for brevity. For $F_{ds}/F_{hg} = 1/10$ the results essentially reproduce the simulations without discrete emission. For $F_{ds}/F_{hg} = 1/3$ the results essentially reproduce the simulations where the gas is rotating; i.e. discrepancy of X-ray data with mass-follows-light assumption at the 68% level, marginal agreement at the 90% level. Higher quality spectral data could better constrain $F_{ds}/F_{hg}$ and thus allow more robust constraints to be obtained.

To give an impression of the combined effects of rotation and discrete sources we considered the rotating model of §5.2.1 with discrete emission of $F_{ds}/F_{hg} = 1/3$. The combined results are similar to what was found for each of the cases separately (see Table 10). For the aperture $a = 75''$ the simulations give $\epsilon_M = 0.12 - 0.18 (0.08 - 0.22)$ at 68%(90%) confidence. It is interesting that the 90% lower limit $\epsilon_M = 0.09$ (i.e. $\epsilon_M^{100}$) for $a = 135''$ is only marginally consistent with the measured value of 0.10; the simulations without rotation or discrete emission have a 90% lower limit of 0.04. Future observations that accurately determine $\epsilon_M$ to large radii will be better able to disentangle the effects of rotation and discrete emission on this geometric test for dark matter.

## 6. Composite Gravitating Matter Distribution

### 6.1. Method

The technique we employ to constrain the shape of the galaxy potential, and hence its mass, from the X-ray image derives from the pioneering work of Binney & Strimple (1978 Strimple & Binney 1979) and is discussed in detail by Buote & Tsai (1995a) and BC94. We refer the reader to these papers for exposition of the modeling procedures we employ in this paper.

We consider the following two families of gravitational potentials:

1. Spheroidal Mass Distributions (SMD): Potentials that are generated by mass distributions stratified on concentric, similar spheroids.

2. Spheroidal Potentials (SP): Potentials that are themselves stratified on concentric, similar spheroids.

Although the SP models have some properties that are undesirable for a physical mass model (i.e. the density may be "peanut-shaped" and possibly somewhere take negative values), the constant shape of the potential and the ellipticity gradient of the mass distribution contrast nicely with the SMDs. The SMD potentials are generated by mass densities $\rho(m)$, where $m^2 = R^2/a^2 + z^2/b^2$, $R$ and $z$ are the conventional cylindrical coordinates, $a$ is the semi-major axis and $b$ the semi-minor axis of the spheroid that bounds the mass; full accounts of SMD potentials are given by Chandrasekhar (1969) and Binney & Tremaine (1987). We consider mass densities having either a Ferrers, $\rho(m) \propto (R_c^2 + a^2 m^2)^{-n}$, or Hernquist (1990), $\rho(m) \propto (am)^{-1}(R_c + am)^{-3}$, form; note that $\rho(m) \equiv 0$ outside of the bounding spheroid. The free parameters of the SMD models are the core parameter, $R_c$, semi-major axis length, $a$, the ellipticity $\epsilon = 1 - b/a$ of the mass, and the power-law index $n$.

The SP models are given by $\Phi = \Phi(\xi)$, where $\xi^2 = R^2 + z^2/q_\Phi^2$; $q_\Phi$ is the constant axial ratio of the SP such that $q_\Phi < 1$ for oblate and $q_\Phi > 1$ for prolate SPs. In particular, we consider the logarithmic potential of Binney (1981; Binney & Tremaine 1987; also Kuijken & Dubinski 1994), $\Phi(R,z) \propto \log(R_c^2 + \xi^2)$, and the power-law potentials (Evans 1994), $\Phi(R,z) \propto (R_c^2 + \xi^2)^{-n}$. The free parameters for these models are $R_c$, $\epsilon_\Phi$ (which is $1 - q_\Phi$ for the oblate models and $1 - 1/q_\Phi$ for prolate models), and $n$ for the power-law models.

Because of the observed elongation of the optical isophotes (see §3) the symmetry axis of the mass of NGC 1332 should not be substantially inclined with respect to the sky plane. That is, even if the outer isophotes ($\epsilon \sim 0.70$) have intrinsic ellipticity appropriate to a galactic disk ($\epsilon \sim 0.90$; Binney & Tremaine 1987), then the galaxy can be inclined at most 15 deg with respect to the sky plane; i.e., $i = 75$ deg. Hence, we do consider the effects of this moderate inclination angle on the estimates of the intrinsic shape of the galaxy. We also consider the effects of gas rotation and emission from discrete sources in the manner discussed previously (§§5.2.1 and 5.2.2).

### 6.2. Shape of the Composite Mass

We determine the intrinsic shape of the underlying galactic mass by comparing $\epsilon_M$ (§2.3.1) and the azimuthally averaged radial profile (§2.2) of the PSPC image to those generated by the models (i.e. model ellipticity $\epsilon_M^{model}$). We quantify the elongation of the model surface brightness within semi-major axes $a = 75''$ and $a = 90''$ because they provide the most stringent constraints on $\epsilon_M$. To save CPU time we use $15''$ pixels in the models for computing the radial profiles; these are then compared to the image radial profile having the $15''$ central bin. However, when comparing ellipticities of select models we always use $5''$ pixels. Note that the uncertainty due to the details of the PSF (see §2.2) affects mostly determination of $R_c$; the slope and ellipticity of the surface brightness, especially for $a = 75'' - 90''$, are hardly affected. The total mass, however, is more sensitive (see §6.3).

For the SMD models we begin by specifying the semi-major axis ($a$) of the bounding mass spheroid, the power-law index of the particular mass model (i.e. Ferrers [$n = 1 - 1.5$] or Hernquist), and the inclination angle ($i$) of the symmetry axis (i.e. either $i = 90\,\mathrm{deg}$ or $i = 75\,\mathrm{deg}$). Then for a given ellipticity of the mass we generate model X-ray surface brightness maps for any values of the free parameters associated with the particular solution of the hydrostatic equation; i.e. $R_c$ and $\Gamma$ for the isothermal case and $R_c$, $\gamma$, and $\Gamma$ for the polytropic case (see BC94). The procedure is the same for the SP models except that (1) the boundary of the mass is not specified, and (2) the power-law index of the potential (logarithmic or $n = 0.1 - 0.5$) and the ellipticity of the potential ($\epsilon_\Phi$) are specified. We determine the confidence limits on the free parameters by performing a $\chi^2$ fit that compares the radial profiles of the model and image. The confidence interval is defined by models having $\chi^2 \leq \chi^2_{min} + \Delta\chi^2$, where $\Delta\chi^2 = (4.61, 6.25)$ for the 90% confidence interval of the isothermal and polytropic cases respectively; $\Delta\chi^2 = (2.30, 3.53)$ and $\Delta\chi^2 = (6.17, 8.02)$ are used for the 68% and 95% confidence limits.

We determine a particular model in these confidence intervals to be consistent with the image if $\epsilon_M$ computed from the model in the $75''$ and $90''$ apertures is consistent with that computed from the image in §2.3.1. The confidence interval used for the radial profile is also used for the $\epsilon_M - \epsilon_M^{model}$ comparison; e.g., 90% confidence limit on $\epsilon_{mass}$ reflects 90% limits on the parameters determined from the radial profile and from $\epsilon_M$.

### 6.2.1. Mass Traces Light

First we examine again in this more detailed analysis if the hypothesis that mass traces light in the galaxy is consistent with the X-ray data. To estimate the luminosity density, $j_L$, we deproject the $I$-band major-axis surface brightness profile (§3). For simplicity we use the single De Vaucouleurs Law ($R_e = 55''$) fitted to the major-axis profile. We assign to $j_L$, and hence to the mass density, $\rho$, the constant intensity-weighted ellipticity of 0.43. Hence our model for the composite mass density, $\rho$, is a Hernquist SMD with $R_c = 55''/1.8153$, $a = 300''$, and $\epsilon_{mass} = 0.43$; the choice for $a$ was arbitrary since any $a \gtrsim 150''$ yields similar major-axis surface brightness profiles.

Assuming the gas is isothermal, the constant $M/L$ model is ruled out to high confidence since it produces an X-ray radial profile that is far steeper than observed; the best-fit yields $\chi^2_{min} = 73$ for 9 degrees of freedom. The polytropic solution, however, is a good fit to the radial profile with best-fit values $\chi^2_{min} = 8.2$ (8 dof), $\gamma = 1.27$, and $\Gamma = 3.79$. Using the same definitions in §4, this model produces a sizeable temperature gradient $\frac{d\ln T_{gas}}{d\ln R} = -0.37$ that is marginally inconsistent with the K-S tests in §4. The ellipticities generated by the polytropic model (and the isothermal model), $\epsilon_M = 0.09(0.10)$ for apertures $a = 75''(90'')$, are inconsistent with the data at 68% confidence and only marginally consistent at 90% confidence; note the effects of rotation like the stars and emission from discrete sources in the galaxy reduce the discrepancy to the 68% confidence level as found in §5.2.1.

Hence, because of the poor constraints on the temperature profile of the gas and the relatively weak constraints on the shapes of the X-ray isophotes, a constant $M/L$ model is inconsistent with the X-ray data at 68% confidence but marginally consistent at the 90% confidence level. The ellipticities generated by the polytropic and isothermal cases are virtually identical and yield results consistent with the geometric test in §5.2. The similarity of the shapes of the isothermal and polytropic models appears to be a generic feature of these models (Strimple & Binney 1979; Fabricant, Rybicki, & Gorenstein 1984; also see appendix B of BC94).

### 6.2.2. General Mass Models

We list in Table 11 the results for the isothermal SMD models; the semi-major axis of the mass models is set to $a = 450''$ which corresponds to $a = 43.6$ kpc for $D = 20$ Mpc; in Figure 9 we display the results for a typical model. The $\rho \sim r^{-2}$ (i.e. $n = 1$) and Hernquist models yield excellent fits to the X-ray radial profile, but the $\rho \sim r^{-3}$ (i.e. $n = 1.5$) model is too steep to fit the data. The derived ranges of $\epsilon_{mass}$ for the $\rho \sim r^{-2}$ and Hernquist models are very similar with $\epsilon_{mass} \gtrsim 0.50$ (68% confidence) marginally inconsistent with the intensity weighted $\epsilon = 0.43$ of the optical isophotes (see §3). At 90% confidence this disagreement vanishes since $\epsilon_{mass} \gtrsim 0.30$. Note that the need for a flattened mass distribution is still significant at the 95% confidence level where, e.g., $\epsilon_{mass} = 0.26 - 0.82$ for the oblate $\rho \sim r^{-2}$ model.

These constraints on $\epsilon_{mass}$ are not sensitive to the value of $a$. There is no upper limit on $a$, but we define a lower limit when models yield $\chi^2_{min}$ such that the probability that $\chi^2$ should exceed $\chi^2_{min}$ is less than 10%. The value of $a_{min}$ slightly varies over the range of $\epsilon_{mass}$. We estimate $a_{min} = 105''$ ($a_{min} = 120''$) at the lower limit for $\epsilon_{mass}$ and $a_{min} = 135''$ ($a_{min} = 135''$) at the upper limit for $\epsilon_{mass}$ for oblate (prolate) models at 90% confidence for the $n = 1$ SMD; there is little difference in $a_{min}$ between the SMD models considered. For $a \geq a_{min}$ the constraints on $\epsilon_{mass}$ systematically differ by $\lesssim 0.02$.

We examined the effects of inclination by setting $i = 75 \deg$ for the isothermal $\rho \sim r^{-2}$ SMD models having $a = 450''$ (see end of §6.1). The 68% and 90% confidence limits for prolate and oblate models are listed in Table 11. The derived $\epsilon_{mass}$ constraints are systematically shifted upwards with respect to the $i = 90 \deg$ models by $\sim 0.06$ for the oblate models and $\sim 0.03$ for the prolate models.

In Table 11 we list the results for the isothermal SP models; in Figure 10 we display the results of a typical model. The logarithmic and $n \lesssim 0.1$ SP models fit the data well, but models with larger values of $n$ do not; e.g., for the $n = 0.25$ model $\chi^2_{min} \sim 18$. Since the SP models have mass distributions that change shape with radius, we assign an aggregate ellipticity to the SP models by computing the ellipticity obtained from the iterative moments (§2.3.1) in a plane containing the major and minor axes. For values of the aperture semi-major axis $\gtrsim 150''$, $\epsilon_{mass}$ is essentially constant because of the small values of $R_c$. In Table 11 we list the average value of $\epsilon_{mass}$ for aperture semi-major axes $\gtrsim 150''$. The logarithmic SP yields $\epsilon_{mass}$ constraints very similar to the oblate SMD models. The prolate SP models unlike the SMD models have $\epsilon_{mass}$ virtually identical to the oblate models; this is simply a manifestation of the different behavior of the SMD and SP models in projection. The power-law models yield very flat masses with no upper limit for $n \gtrsim 0.1$.

The polytropic models also require flattening of the underlying mass. We obtain similar results for the $\rho \sim r^{-2}$ SMD polytropic model; i.e. at 90% confidence limits $\epsilon_{mass} = 0.33 - 0.80 (0.30 - 0.71)$ and $\gamma = 0.73 - 1.38 (0.76 - 1.41)$ for oblate (prolate) models with $a = 450''$. The oblate models have corresponding temperature gradients $\left|\frac{d \ln T_{gas}}{d \ln R}\right| \leq 0.39$ which are essentially consistent with the PSPC spectrum (see §4); i.e. even with the additional freedom allowed by temperature gradients, the constraints on $\epsilon_{mass}$ are essentially identical to the isothermal case for the $\rho \sim r^{-2}$ SMD model. The polytropic $\rho \sim r^{-3}$ SMD model fits the surface brightness well but requires large polytropic indices $\gamma = 1.24 - 1.89$ (oblate) some of which are unlikely to be present in early-type galaxies. The oblate (prolate) models give a 90% lower limit $\epsilon_{mass} \geq 0.53 (0.43)$. For each acceptable $\rho \sim r^{-3}$ SMD model the 90% lower-limit $\gamma \sim 1.25$ implies $\left|\frac{d \ln T_{gas}}{d \ln R}\right| \approx -0.34$ which is only marginally consistent with the spectral constraints in §4. The Hernquist polytropic models behave like the $\rho \sim r^{-2}$ models in that they reproduce the isothermal results for $\epsilon_{mass}$, $R_c$, and $\Gamma$ in the 90% confidence interval, but the best-fit region of the parameter space has unphysical parameters such as huge values of $R_c \gg a$ and small values of $\gamma < 0.4$; these low values of $\gamma$ typically imply $\left|\frac{d \ln T_{gas}}{d \ln R}\right| \gtrsim 0.7$ which is inconsistent with the spectral constraints in §4. The logarithmic SP model yields $\epsilon_{mass} \geq 0.36$ (90% confidence) with $R_c$ and $\Gamma$ values comparable to their isothermal ranges, although no upper limit can be set on $\epsilon_{mass}$ at 90% confidence. Thus, the presence of possible temperature gradients does not greatly affect the constraints on the intrinsic shape of NGC 1332; even the weak spectral constraints obtained from the PSPC data (§4) strongly suggest that $\rho$ is flatter than $r^{-3}$ for the SMD models.

We examined the effects of rotation following our treatment in §5.2.1. In Table 11 we list the results for the oblate Hernquist models ($a = 450''$) having rotation similar to that of the stars. The derived ellipticity ranges are systematically shifted down by $\Delta \epsilon_{mass} \sim 0.15$. However, the need for a flattened halo is still preserved even considering the substantial rotation.

We estimate the effect of emission from discrete sources on $\epsilon_{mass}$. In Table 11 we list the results for the isothermal oblate $\rho \sim r^{-2}$ SMD model of the hot gas with a discrete model (see §5.2.2) in terms for their flux ratio $F_{ds}/F_{hg}$. Models with $F_{ds}/F_{hg} = 1/2$ (i.e. 1/3 discrete emission, 2/3 hot gas) have the largest discrete contribution that yields acceptable fits to the data; e.g., for $F_{ds}/F_{hg} = 1$, $\chi^2_{min} \sim 49$ for 8 dof – unacceptable models have $\chi^2_{min}$ such that the probability that $\chi^2$ should exceed $\chi^2_{min}$ is less than 10%; note also that $F_{ds}/F_{hg} \lesssim 1$ implies discrete fluxes $F_{ds}$ consistent with the $L_X/L_B$ relation for galaxies dominated by emission from discrete sources (see, e.g., Canizares et al. 1987). The upper limit on $\epsilon_{mass}$ is virtually unaffected by the presence of discrete emission, but the lower limit is very sensitive; e.g., the lower limit on $\epsilon_{mass}$ falls to 0.16 (90% confidence) for $F_{ds}/F_{hg} = 1/2$. Hence, the ability to constrain the shape of the underlying mass is substantially influenced by the possible presence of emission from discrete sources in the galaxy.

Finally, we considered the combined effects of rotation and emission of discrete sources using the rotating model of §5.2.1 with a discrete component such that $F_{ds}/F_{hg} = 1/3$ in analogy to what was done in §5.2.2; the results are presented in Table 11. As expected, the effects of both rotation and discrete emission further weaken, but do not eliminate, the need for a substantially flattened halo.

### 6.3. Estimate of the Composite Mass Profile

Using the models of the previous section together with the results for the PSPC spectrum (§4) we compute the total integrated mass of the galaxy; note that (unlike the shape) the procedure to obtain the total mass profile (see, e.g., eq. [26] of BC94) *requires detailed knowledge of the temperature profile and thus suffers from the same uncertainties of the traditional spherically-symmetric analysis of the total mass* (for a review see Fabbiano 1989). Some examples of pre-ROSAT studies that exhibited X-ray evidence for dark matter in early-type galaxies are Fabian et al. (1986), who estimated a lower limit for the total binding mass of a sample of early-type galaxies using IPC data in a manner free from detailed knowledge of the gas temperature profiles, Loewenstein & Mathews (1987), who studied evolutionary models of gaseous halos in early-type galaxies, and Loewenstein (1992), who analyzed combined optical and IPC X-ray data for the Virgo galaxy NGC 4472.

In Table 12 we list the total masses ($M_{tot}$) and total blue mass-to-light ratios ($\Upsilon_B = M_{tot}/L_B$) for several mass models of the previous section. The total blue luminosity, $L_B = 1.91 \times 10^{10} L_\odot$, is taken from Donnelly et al. (1990) scaled to $D = 20$ Mpc. The 68% and 90% confidence limits reflect the same confidence values for the parameters of the mass models. However, in both cases the 90% confidence limits on the gas temperature are used. Note that the uncertainty in $R_c$ due to PSF calibration (§2.2) is less than 50% which typically translates to at most 30% uncertainty in the mass.

The total masses overlap for the all of the models although there are systematic differences between models; e.g., $n = 1$ isothermal SMD models systematically allow for higher masses than the steeper isothermal Hernquist models. The lower bound on $\Upsilon_B$ is obtained from the $a_{min}$ prolate model which gives $\Upsilon_B = 8.9 \Upsilon_\odot$ at 68% confidence and $\Upsilon_B = 7.9 \Upsilon_\odot$ at 90% confidence (assuming $D = 20$ Mpc) for the isothermal $n = 1$ SMD models. The 90% lower limits on $\Upsilon_B$ are not very different from $\Upsilon_B \sim 7 \Upsilon_\odot$ expected of the stellar population alone (§7.1). The effects of possible temperature gradients are substantial for the total mass and yield mass ranges larger than a factor of two over the isothermal case. In Figure 11 we plot the integrated mass profile as a function of spheroidal radius $am$, where $a$ is the semi-major axis of the spheroid having mass $M_{tot}$, for two typical SMD models consistent with the fits to the radial profile; we also plot the spherically-averaged mass profile for the isothermal logarithmic SP model. Assuming $\Upsilon_B \sim 7\Upsilon_\odot$ for the stellar population and that the gas mass may be neglected (§7.2) we compute the relative fraction of mass in stars ($M_{stars}$) and dark matter ($M_{DM}$). For the isothermal $\rho \sim r^{-2}$ SMD models we obtain 90% confidence estimates of $M_{DM}/M_{stars} = 2.9 - 11.1$ for $a = 450''$ and $M_{DM}/M_{stars} = 1.4 - 3.3$ for the minimum acceptable $a$ models.

We examined the effects of gas rotation and emission from discrete sources on determining the integrated mass profile using a series of isothermal Hernquist models listed in Table 12: (1) Hernquist SMD ($a = 450''$), (2) rotation model, (3) Hernquist SMD with discrete component in flux ratio $F_{ds}/F_{hg} = 1/3$, and (4) rotation model plus discrete component in flux ratio $F_{ds}/F_{hg} = 1/3$. The total masses for these models agree to within 30%. The ellipticity constraints for these models, in contrast, vary substantially more than this (see Table 11); e.g., the 90% confidence limits $\epsilon_{mass} = 0.1 - 0.6$ for model (4) and $\epsilon_{mass} = 0.33 - 0.83$ for (1) corresponding to a 40% shift in the mean for the two models. Hence, the X-ray determination of the integrated mass profiles of early-type galaxies are less sensitive to uncertainties due to possible rotation of the gas and emission from discrete sources than the intrinsic shape, $\epsilon_{mass}$.

Finally, we compute the circular velocity ($v_c$) of these models. Franx (1993) shows that simple models of ellipticals with massive halos satisfy a Tully-Fisher relation when $v_c/\sigma_0 \sim 1.38$, where $v_c$ is the maximum circular velocity of the halo and $\sigma_0$ is the observed central velocity dispersion. Dalle Ore et al. (1991) measure $\sigma_0 = 347$ km s$^{-1}$ for NGC 1332 which has the distinction of having the largest $\sigma_0$ in their sample of 79 early-type galaxies. For the $\rho \sim r^{-2}$ SMD models, $v_c$ occurs at $a$ which gives 90% confidence values of $v_c = 278 - 428$ km s$^{-1}$ and $v_c/\sigma_0 = 0.80 - 1.23$ for $a = 450''$; at the minimum values of $a$, $v_c = 313 - 453$ km s$^{-1}$ (90% confidence) and $v_c/\sigma_0 = 0.90 - 1.31$. The logarithmic SP model gives 90% confidence values $v_c = 272 - 414$ km s$^{-1}$ and $v_c/\sigma_0 = 0.78 - 1.20$. These values for $v_c/\sigma_0$ are systematically less than the suggested value from Franx (1993), but the large central velocity dispersion for NGC 1332 may suggest that it is an unusual case.

## 7. Dark Matter Distribution

The composite gravitational potential we analyzed in the previous section may be decomposed into its constituents,

$$\Phi = \Phi_{stars} + \Phi_{gas} + \Phi_{DM}, \tag{5}$$

where $\Phi_{stars}$ is the potential arising from a mass distributed like the light having the total mass of the visible stars, $\Phi_{gas}$ is the potential generated by the the mass of the X-ray–emitting gas, and $\Phi_{DM}$ is the potential generated by the dark matter. By using the observations to constrain $\Phi_{stars}$ and $\Phi_{gas}$ we may determine the distribution of dark matter directly. The procedure to constrain $\Phi$ is the same as before except now the free parameters are those associated with the shape of $\Phi_{DM}$ and the ratio of the mass in dark matter to the mass in visible stars, $M_{DM}/M_{stars}$ (see BC94); note we neglect $\Phi_{gas}$ since $M_{gas} \ll M_{stars}$ (see below).

BC94 pointed out that determination of $M_{DM}/M_{stars}$ in this fashion depends only on the relative distribution of stars and dark matter, not the distance to the galaxy. As a result, BC94 suggested (but did not explain) that in principle $M_{DM}/M_{stars}$ could be used a distance indicator. If the stellar mass-to-light ratio, $\Upsilon_{stars}$, of early-type galaxies is universal, then $M_{stars} = \Upsilon_{stars} L_{stars} \propto D^2$, where $D$ is the distance to the galaxy. For $M_{DM}/M_{stars}$ constrained by fitting to the surface brightness, we may compute the resulting total mass $M_{fit} = M_{stars} + M_{DM} \propto D^2$. Comparing $M_{fit}$ to $M_{tot} \propto D$ computed in §6.3, we have $D \propto M_{fit}/M_{tot}$. Hence for high-quality optical and X-ray data that accurately determine $M_{fit}$ and $M_{tot}$ the distance to the galaxy may in principle be inferred. Unfortunately, the value of $\Upsilon_{stars}$ is at present quite uncertain (see below). In any event, spatial and spectral X-ray data far superior to the PSPC would be required to accurately address, e.g., the issues of temperature gradients, rotation, emission from discrete sources and thus enable examination of the viability of this technique as a distance indicator.

### 7.1. Visible Stellar Mass Distribution

The stellar distribution is just that of the constant $M/L$ model of §6.2.1. Following BC94 we normalized $\rho_{stars}$ by assigning the stars a mass-to-light ratio $\Upsilon_B = 7\Upsilon_\odot$. This value is actually quite uncertain since population synthesis studies give $\Upsilon_B = (1 - 12)\Upsilon_\odot$ (e.g., Pickels 1985). The uncertainty is reduced by normalizing to dynamical studies of the cores of ellipticals (e.g., Bacon 1985; Peletier 1989). As a result, $\Upsilon_B = 7\Upsilon_\odot$ may overestimate the stellar mass by including a significant amount dark matter. With these uncertainties aside, we obtain a stellar mass $M_{stars} = 1.3 \times 10^{11} M_\odot$ assuming $D = 20$ Mpc.

### 7.2. X-ray Gas Mass Distribution

The radial X-ray surface brightness distribution is well parametrized by the $\beta$-model (§2.2), but the elongation is not so well determined (§2.3.1). We compute the radial distribution under the assumption of spherical symmetry which should not introduce errors greater than $\sim 25\%$ for plausible ellipticities of the gas (e.g., BC94). Straightforward deprojection of the $\beta$-model yields the volume emissivity, $j_{gas}$. The gas density, $\rho_{gas} \propto \sqrt{j_{gas}}$ (see BC94), is then integrated to get the total gas mass.

We list in Table 13 the gas mass, $M_{gas}$, the volume-averaged particle density, $\bar{n}$, and its associated cooling time, $\bar{\tau}$, all computed within the gas sphere with radius $r = 300''$; note these values reflect 90% confidence values on $a_X$, $\beta$, and the spectral parameters of the single-temperature model (§4). The 90% confidence range of $M_{gas}$ implies that $M_{gas}/M_{stars} \lesssim 1/100$. For such small values the self-gravitation of the gas may be neglected in eq. [5] and hence we ignore $\Phi_{gas}$ in the following determination of the dark matter.

### 7.3. Results

For a given value of $M_{DM}/M_{stars}$ the fitting procedure follows that in §6.2 where the free parameters are now associated with the dark matter model. Because adding another parameter weakens the already loose constraints on the models, we restrict our discussion to isothermal $\rho \sim r^{-2}$ SMD models for the dark matter. Similarly, we do not discuss the effects of rotation or emission from discrete sources here. We simply wish to obtain an understanding of how the dark matter itself is distributed for some plausible models.

In Table 14 we list the results for several values of $M_{DM}/M_{stars}$ where the semi-major axis of the dark matter has been set to $a = 450''$; note the quality of the fits and isophote shapes for typical models are the same as for the composite case (§6.2). For $M_{DM}/M_{stars} > 10$ the dark matter models agree well with the results for the composite mass in §6.2. Smaller values of $M_{DM}/M_{stars}$ render the models less constrained. In particular, the upper limit on $\epsilon_{DM}$ becomes indeterminate while the lower limit holds firm at about 0.31 (90% confidence). Employing the same criteria of goodness of fit used to constrain $a$ in §6.2, we determine a lower limit $M_{DM}/M_{stars} > 3$ for these models having $a = 450''$; the results are not sensitive to large values of $a$. These values of $M_{DM}/M_{stars}$ are consistent to those obtained using the gas temperature in §6.3.

We are unable to set a lower limit on $a$ for the dark matter models because of the extra freedom given by the choice of $M_{DM}/M_{stars}$. However, for $M_{DM}/M_{stars}$ as small as 2, the lower limit on $a$ is essentially the same as that for the composite mass in §6.2 and the shapes are essentially the same as the $a = 450''$ case to within $\sim 10\%$. For smaller $M_{DM}/M_{stars}$ the results are highly uncertain and are, to a large extent, much more sensitive to the precise form of the model for $\Phi_{stars}$ than the models with larger $M_{DM}/M_{stars}$ values.

In Figure 12 we plot the major-axis mass profiles of the stars, gas, and dark matter for a typical model having $M_{DM}/M_{stars} = 10$; the plot is normalized to the value of $M_{stars}$ (§7.1). We also plot in Figure 12 the spatially variant $\Upsilon_B$ along the major axis where $\Upsilon_B(ma) = [M_{stars}(ma) + M_{DM}(ma)]/L_B(ma)$, $L_B(ma) = \Upsilon_B^{stars} M_{stars}(ma)$, and $\Upsilon_B^{stars} = 7\Upsilon_\odot$ as given in §7.1. At $\sim 10''$, the stellar and dark mass contribute equally. The dark matter rises faster than the stars and dominates the total mass exterior to $R_e$. This description is similar to what was found for NGC 720 by BC94.

## 8. Discussion

The geometric test for dark matter introduced by BC94 and clarified in this paper (§5.1) is the most robust means to test the hypothesis of mass tracing light for flattened, X-ray–bright, early-type galaxies. Optical methods suffer from uncertainty in the shape of the velocity dispersion tensor and thus generally test the mass-follows-light hypothesis by assuming the phase-space distribution function depends only on two integrals of motion (e.g, Binney et al. 1990; van der Marel 1991). Standard X-ray analyses are hindered by poorly determined temperature profiles of the gas (e.g., Fabbiano 1989). Even when the temperature profile could in principle be determined from high quality spatially resolved spectra, interpretation of the spectra in terms of a temperature profile is still uncertain due to questions about the reliability of standard spectral models (Trinchieri et al. 1994). The geometric test only requires knowledge of $\Sigma_X$ and a few additional unrestrictive assumptions (see §5.1). Although ROSAT has accurate maps of $\Sigma_X$ for only a few galaxies, many more should become available with the superior resolution of the Advanced X-ray Astrophysics Facility (AXAF).

It is important to emphasize that the detailed analysis of the mass distribution in §6 and §7 inherently depends to some extent on the models used. Only with detailed two-dimensional temperature and surface brightness maps can the potential be usefully determined without parametrizations for $\Phi$ and $T_{gas}$. To determine robustly whether the mass models of §6 and §7 are consistent with the X-ray data independent of the poorly constrained temperature distribution, we could use the geometric test discussed in §5.1 by replacing $\Phi_L$ with the potential derived from the mass models. This approach has the disadvantage that only one model (set of parameters) may be tested at a time.

The relatively poor constraints on the ellipticity of the X-ray isophotes (§2.3.1) are primarily responsible for the large range of acceptable $\epsilon_{mass}$. In Figure 13 (a) we plot $\epsilon_M$ as a function of aperture semi-major axis for six typical oblate isothermal models of the composite mass; also plotted are the 90% confidence intervals on $\epsilon_M$ computed from the PSPC data (§2.3.1). All of the models produce very similar $\epsilon_M$ for $a \lesssim 100''$ where the data is best determined. At larger radii, where the discrepancy between models is most pronounced, $\epsilon_M$ is poorly constrained. However, in Figure 13 (b) we plot the true expected isophotal ellipticity of the models; i.e. the ellipticity for a given isophote assuming a very narrow PSF. Now we see that the $\epsilon$ profiles are substantially different for the models. For example, the $\epsilon$ profile of the Hernquist SMD model falls more steeply than the other models, but adding rotation does flatten it out somewhat; the $n = 1$ SMD and logarithmic SP models have nearly identical profiles outside $R_e$, but inside the SMD profile increases while that of the SP remains constant. Moreover, the models with a discrete component have a bump in the profiles near $R_e$ not seen in the other models. Hence, isophotal ellipticities computed to better than $\Delta \epsilon \sim 0.02$ over the range $10'' - 200''$ would serve as a powerful discriminator for the mass of the galaxy.

Although the shape of the dark halo in NGC 1332 is uncertain, i.e. ranging from $\epsilon_{mass} = 0.12 - 1$ considering all of the models (90% confidence), if rotation of the gas and emission from discrete sources are indeed negligible then the models suggest $\epsilon_{mass} \sim 0.5 - 0.6$. This degree of flattening for the dark halo is consistent with NGC 720 (BC94) and appears consistent with recent studies of the shapes of dark halos; see Sackett et al. 1994 for a discussion. The flattening is also consistent with results from N-body simulations of dissipationless collapse incorporating the effects of a small dissipational component (Katz & Gunn 1991; Dubinski 1994).

## 9. Conclusion

We have analyzed ROSAT PSPC X-ray data of the S0 galaxy NGC 1332 for the purpose of constraining the intrinsic shape of its underlying mass and presenting a detailed investigation of the uncertainties resulting from the assumptions underlying this type of analysis. Our treatment closely parallels that of BC94 who analyzed the shape of the elliptical galaxy NGC 720 using ROSAT PSPC data. NGC 1332 was selected because, along with NGC 720, it has among the largest known X-ray fluxes from *Einstein* (Fabbiano et. al. 1992) for an early-type galaxy that is elongated in the optical ($\epsilon \sim 0.43$), relatively isolated from other large galaxies, has an angular size ($\sim 7'$) substantially larger than the PSPC PSF, and is likely to be dominated by emission from hot gas (Kim et al. 1992).

The ellipticity of the surface brightness is computed by taking quadrupole moments of the X-ray surface brightness in elliptical apertures of different semi-major axis length. The isophote shapes are best constrained for semi-major axes $a = 75'' - 90''$ where $\epsilon_M = 0.10 - 0.27$ (90% confidence). The position angles of the X-ray isophotes for different $a$ are consistent with each other and the optical value of 115 deg N-E within the estimated 95% uncertainties.

The spectrum is not well constrained by the PSPC data. A single-temperature plasma with Galactic column density fits the data well, but the temperature and metallicities are not well determined; e.g., $T_{gas} = 0.40 - 0.76$ keV and abundances = $10\% - 107\%$ solar (90% confidence). Radial temperature gradients are not required by the PSPC spectrum, but large gradients $\left| \frac{d \ln T_{gas}}{d \ln R} \right| < 0.35$ (99% confidence) are consistent with the spectrum. Although simple models for emission from discrete sources in NGC 1332 can be ruled out as the sole source of the X-ray emission, the PSPC spectrum does not tightly constrain the relative flux of hot gas and discrete emission; generally as high as equal fluxes in the two components are allowed by the spectrum.

We clarified the geometric test for dark matter introduced by BC94. Besides a few unrestrictive assumptions, this test allows in principle a model-independent test for whether mass follows light in the galaxy by making the fundamental assumption that the X-rays result from hot gas in hydrostatic equilibrium with the gravitational potential of the galaxy. By applying a version of this test suitable for the low spatial resolution and low $S/N$ PSPC data for NGC 1332, we find that mass tracing the optical light is not consistent with the X-ray data at the 68% confidence level, but marginally consistent at the 90% level. We considered both the effects of possible rotation of the gas and emission from discrete sources (which are not important for NGC 720 – see BC94). Following BC94 we assert that the same test may be directly applied to MOND with the same implications, although with weaker significance than NGC 720.

Explicitly solving the equation of hydrostatic equilibrium, we analyzed the shape and profile of the mass using the technique of BC94 (also Buote & Tsai 1995a) derived from the work of Binney & Strimple (1978; Strimple & Binney 1979). By employing a wide class of spheroidal models we constrained the ellipticity of the underlying mass to be $\epsilon_{mass} \sim 0.5 - 0.7 (0.3 - 0.8)$ at 68% (90%) confidence assuming the rotation of the gas and emission from discrete sources are negligible; a model where the mass traces the optical light is easily ruled out for an isothermal gas, but polytropic models, though inconsistent at the 68% confidence level, are marginally consistent with the data at 90% confidence. Considering the possible effects of rotation of the gas like that of the stars and the effects of emission from discrete sources, the constraints on $\epsilon_{mass}$ are weakened considerably. For all the isothermal models considered, the 90% confidence estimate of the total mass out to $a = 450''$ is $M_{tot} = (0.38 - 1.7) \times 10^{12} M_\odot$ assuming $D = 20$ Mpc corresponding to blue mass-to-light ratio $\Upsilon_B = (31.9 - 143) \Upsilon_\odot$.

We estimated the observed stellar and X-ray mass profiles from their observed surface brightness distributions and then fit the dark matter directly. For models where $M_{DM}/M_{stars} > 10$, the dark matter has the same shape as for the above composite models. For smaller values of $M_{DM}/M_{stars}$ the lower limit on $\epsilon_{mass}$ falls by 0.03 but the upper limit is indeterminate (90% confidence). We estimate a lower limit $M_{DM}/M_{stars} > 3$ (90% confidence) assuming the dark matter extends out to $a = 450''$; these results are not sensitive to $a \gtrsim 250''$.

In all, there is marginal evidence for a flattened halo ($\epsilon \approx 0.3 - 0.8$, 90% confidence) of dark matter in NGC 1332 consistent with the elongated halo of NGC 720 ($\epsilon \approx 0.5 - 0.7$, 90% confidence) we found in BC94. The X-ray analysis of NGC 720 and NGC 1332 suggest that the dark matter halos in early-type galaxies are indeed substantially flattened (as also suggested by the analysis of the polar-ring galaxy NGC 4650A by Sackett et al. 1994) in agreement with the predictions of N-body/hydrodynamic simulations (e.g., Dubinski 1994).

We have demonstrated the need for obtaining higher quality spatial and spectral X-ray data to place rigorous constraints on the shape and amount of mass in early-type galaxies. Specifically, (1) the shapes of the X-ray isophotes from the center out to a few optical $R_e$, (2) the temperature profile of the gas, (3) the flux resulting from discrete sources in the galaxy, and (4) the rotation curve of the gas need to be measured accurately before a truly robust measurement of the shape of the underlying mass can be realized. Although there is no astrophysical instrument on the horizon that will be able to detect rotation of the gas in early-type galaxies, AXAF will greatly improve our understanding of issues (1) - (3).

We acknowledge useful discussions with Eric Gaidos. DAB thanks Eugene Magnier for tutelage in the fine art of optical observing and for providing the flat fields. We thank John

Tonry for providing deep *I*-band surface photometry and a SBF distance for NGC 1332. We appreciate the help of the friendly people at hotseat@cfa.harvard.edu (especially Kathy Manning and Janet De Ponte) for answering questions about PROS and Gail Rohrbach at GSFC for her patient elucidation of the arcane lore of PROS/XSPEC file conversion. We acknowledge use of the SIMBAD data base and the ADS abstract service. Supported in part by grants NAS8-38249 and NASGW-2681 (through subcontract SVSV2-62002 from the Smithsonian Astrophysical Observatory).

Table 1: ROSAT Observation of NGC 1332

| ROSAT Sequence Number | Date Observed | R.A., Decl.[a] | R.A., Decl.[b] | Exposure Time (s) | Flux[c] (erg cm$^{-2}$ s$^{-1}$) |
|---|---|---|---|---|---|
| rp600006 | Aug 13-14, 1991 | $3^h 26^m 17^s, -21 \deg 20' 09''$ | $3^h 26^m 17^s, -21 \deg 19' 57''$ | 25,637s | $4.4(3.8) \times 10^{-13}$ |

[a] Optical center (J2000) from de Vaucouleurs et al. (1991).

[b] X-ray centroid (J2000) computed in this paper.

[c] Computed in 120″ radius circle for energy range 0.1 - 2.4 (0.4 - 2.4) keV.

Table 2: Identified Point Sources

| Source # | R.A. | Decl. |
|---|---|---|
| 1 | $3^h 26^m 14^s$ | $-21 \deg 21' 10''$ |
| 2 | $3^h 26^m 06^s$ | $-21 \deg 20' 03''$ |
| 3 | $3^h 26^m 02^s$ | $-21 \deg 17' 14''$ |
| 4 | $3^h 26^m 36^s$ | $-21 \deg 17' 58''$ |
| 5 | $3^h 26^m 43^s$ | $-21 \deg 15' 19''$ |

Note. — These sources (expressed in J2000 coordinates) were identified and excluded from analysis by either symmetric substitution (§2) or masking (§4).

Table 3: Fits to the $\beta$ Model

| | Best Fit $a_x$ (arcsec) | 90% Range | Best Fit $\beta$ | 90% Range | $\chi^2_{min}$ | Degrees of Freedom | R (arcsec) |
|---|---|---|---|---|---|---|---|
| (1) | 0.1 | < 1.8 | 0.47 | 0.45 - 0.50 | 15.3 | 10 | 285 |
| (2) | 0.8 | 0.004 - 3.4 | 0.47 | 0.44 - 0.50 | 10.2 | 8 | 285 |

Note. — The results of fitting the X-ray radial profile (0.4 - 2.4 keV) to the $\beta$-model (§2.2) for the inner 15″ consisting of (1) three 5″ bins and (2) 1 bin.

Table 4: X-ray Ellipticities

| $a$ (arcsec) | $\epsilon_M$ | 68% | 90% | 95% | 99% | Counts | $\epsilon_{iso}$ | $\Delta\epsilon_{iso}$ |
|---|---|---|---|---|---|---|---|---|
| 30 | 0.12 | 0.06 - 0.15 | 0.00 - 0.22 | 0.00 - 0.25 | 0.00 - 0.30 | 562 | 0.01 | 0.08 |
| 45 | 0.17 | 0.15 - 0.21 | 0.08 - 0.26 | 0.03 - 0.28 | 0.00 - 0.32 | 694 | 0.13 | 0.08 |
| 60 | 0.13 | 0.08 - 0.15 | 0.00 - 0.20 | 0.00 - 0.22 | 0.00 - 0.26 | 767 | 0.26 | 0.20 |
| 75 | 0.20 | 0.16 - 0.22 | 0.10 - 0.27 | 0.08 - 0.28 | 0.00 - 0.32 | 824 | 0.27 | 0.10 |
| 90 | 0.19 | 0.16 - 0.22 | 0.10 - 0.27 | 0.08 - 0.28 | 0.00 - 0.32 | 869 | 0.26 | 0.15 |
| 105 | 0.14 | 0.10 - 0.16 | 0.00 - 0.22 | 0.00 - 0.24 | 0.00 - 0.28 | 922 | | |
| 120 | 0.13 | 0.08 - 0.15 | 0.00 - 0.22 | 0.00 - 0.24 | 0.00 - 0.28 | 953 | | |
| 135 | 0.10 | 0.00 - 0.11 | 0.00 - 0.18 | 0.00 - 0.22 | 0.00 - 0.26 | 968 | | |

Note. — The values of $\epsilon_M$ (and confidence limits) are computed within an aperture of semi-major axis $a$ on the image having $5''$ pixels (§2.3.1) with the background included; the counts, however, have the background subtracted. The results from isophote fitting, $\epsilon_{iso}$, are computed on the image with $15''$ pixels and the uncertainty, $\Delta\epsilon_{iso}$, reflects 68% confidence error estimates.

Table 5: X-ray Position Angles (N through E)

| $a$ (arcsec) | $\theta_M$ | 68% | 90% | 95% | 99% | $\theta_{iso}$ | $\Delta\theta_{iso}$ |
|---|---|---|---|---|---|---|---|
| 30 | 161 | 138 - 183 | 120 - 200 | 107 - 208 | 82 - 235 | 177 | 237 |
| 45 | 132 | 120 - 147 | 111 - 152 | 107 - 157 | 91 - 167 | 137 | 18 |
| 60 | 123 | 106 - 139 | 92 - 155 | 78 - 166 | 53 - 190 | 116 | 23 |
| 75 | 122 | 113 - 132 | 106 - 138 | 104 - 141 | 96 - 150 | 111 | 12 |
| 90 | 111 | 111 - 121 | 95 - 128 | 92 - 133 | 83 - 147 | 99 | 20 |
| 105 | 91 | 78 - 106 | 66 - 117 | 58 - 126 | 35 - 154 | | |
| 120 | 106 | 88 - 124 | 68 - 140 | 52 - 156 | 28 - 187 | | |
| 135 | 118 | 95 - 144 | 71 - 167 | 55 - 185 | 31 - 202 | | |

Note. — Position angles are prepared in the same manner as the ellipticities in Table 4.

Table 6: $I$-Band Ellipticities and Position Angles

| | MDM 1.3m | | | | | Tonry | |
|---|---|---|---|---|---|---|---|
| $a$ (arcsec) | $\epsilon$ | $\Delta\epsilon$ | $\theta$ | $\Delta\theta$ | $a$ (arcsec) | $\epsilon$ | $\theta$ |
| 2.0 | 0.134 | 0.012 | 105.6 | 2.8 | 12.9 | 0.354 | 116.6 |
| 2.4 | 0.164 | 0.010 | 109.6 | 1.9 | 15.5 | 0.393 | 116.6 |
| 2.9 | 0.195 | 0.007 | 111.1 | 1.2 | 18.7 | 0.437 | 116.5 |
| 3.5 | 0.216 | 0.006 | 112.8 | 0.9 | 22.2 | 0.487 | 116.5 |
| 4.3 | 0.230 | 0.004 | 114.1 | 0.6 | 26.3 | 0.582 | 115.8 |
| 5.2 | 0.240 | 0.005 | 114.7 | 0.7 | 30.9 | 0.621 | 115.3 |
| 6.3 | 0.253 | 0.005 | 114.9 | 0.6 | 36.0 | 0.617 | 115.3 |
| 7.6 | 0.274 | 0.005 | 115.8 | 0.7 | 41.8 | 0.646 | 115.0 |
| 9.2 | 0.296 | 0.007 | 116.1 | 0.8 | 48.2 | 0.668 | 114.9 |
| 11.1 | 0.320 | 0.007 | 116.2 | 0.8 | 55.4 | 0.679 | 114.7 |
| 13.4 | 0.350 | 0.009 | 116.5 | 0.9 | 63.3 | 0.683 | 114.7 |
| 16.2 | 0.390 | 0.013 | 116.5 | 1.1 | 72.1 | 0.681 | 114.6 |
| 19.7 | 0.432 | 0.016 | 116.3 | 1.4 | 81.7 | 0.673 | 114.6 |
| 23.8 | 0.478 | 0.024 | 116.4 | 1.9 | 92.3 | 0.659 | 114.5 |
| 28.8 | 0.529 | 0.027 | 115.6 | 2.0 | 103.8 | 0.640 | 114.3 |
| 34.8 | 0.577 | 0.026 | 115.3 | 1.9 | 116.4 | 0.615 | 114.1 |
| 42.2 | 0.626 | 0.029 | 114.8 | 1.9 | 130.1 | 0.580 | 113.4 |
| 51.0 | 0.654 | 0.019 | 115.0 | 1.2 | 145.1 | 0.545 | 112.9 |
| 61.7 | 0.682 | 0.018 | 114.7 | 1.1 | 161.2 | 0.505 | 111.6 |
| 74.7 | 0.688 | 0.026 | 114.8 | 1.6 | 178.7 | 0.473 | 110.1 |

Table 7: Spectral Data and Fit Results

| Region | Model | $\chi^2_{min}$ | dof* | $N_H$ ($10^{20}$ cm$^{-2}$) | Abundance (% solar) | $T$ (keV) |
|---|---|---|---|---|---|---|
| 0″- 120″ | 1T[†] | 9.2 | 11 | (0.083 − 4.7) | 10 - 107 | 0.40 - 0.76 |
| 0″- 30″ | 1T | 2.1 | 4 | (0.13 − 6.0) | > 21 | 0.39 - 0.72 |
| 30″- 120″ | 1T | 2.3 | 4 | < 4.8 | ... | 0.32 - 1.0 |
| (A) | 1T | 2.6 | 4 | < 4.8 | > 44 | 0.36 - 0.77 |
| (B) | 1T | 2.3 | 4 | < 5.7 | 0.4 - 203 | 0.38 - 0.84 |
| 0″- 120″ | (1)[‡] | 34.0 | 12 | | | |
| 0″- 120″ | (2) | 27.6 | 12 | | | |
| 0″- 120″ | (3) | 66.7 | 12 | | | |

---

[*]Degrees of freedom

[†] Single-temperature Raymond-Smith model. 90% confidence estimates for parameters are shown for 0″- 120″, 68% confidence for the others.

[‡]Numbers correspond to (1) power law, (2) $T = 8$ keV Bremsstrahlung, and (3) $T = 8$ keV Raymond-Smith plasma models all with interstellar absorption

Table 8: Qualitative Predictions of Mass-Traces-Light

| $a$ (arcsec) | $\epsilon_\Phi^{3d}$ | $\epsilon_\Phi^{2d}$ | $\langle\epsilon_\Phi^{2d}\rangle_{PSPC}$ | $\epsilon_{isothermal}$ |
|---|---|---|---|---|
| 30 | 0.17 | 0.15 | 0.13 | 0.09 |
| 45 | 0.15 | 0.13 | 0.12 | 0.11 |
| 60 | 0.13 | 0.12 | 0.10 | 0.11 |
| 75 | 0.12 | 0.11 | 0.09 | 0.11 |
| 90 | 0.11 | 0.10 | 0.09 | 0.10 |
| 105 | 0.10 | 0.10 | 0.08 | 0.09 |
| 120 | 0.09 | 0.09 | 0.07 | 0.08 |
| 135 | 0.09 | 0.08 | 0.07 | 0.08 |

Note. — Ellipticities for the oblate mass-traces-light model (see §5.2) of the three-dimensional isopotential surfaces ($\epsilon_\Phi^{3d}$), the contours of constant projected potential ($\epsilon_\Phi^{2d}$), $\epsilon_\Phi^{2d}$ convolved with the PSPC PSF ($\langle\epsilon_\Phi^{2d}\rangle_{PSPC}$), and the predicted X-ray isophotes assuming an isothermal ideal gas (see §6.2.1).

Table 9: Geometric Test for Dark Matter - $\epsilon_M^{max}$ For Simple Case

| $a$ (arcsec) | $\epsilon_X$ | Oblate $\epsilon_M^{max}$ | | Prolate $\epsilon_M^{max}$ | |
|---|---|---|---|---|---|
| | | 68% | 90% | 68% | 90% |
| 30 | 0.12 | 0.09 | 0.13 | 0.10 | 0.13 |
| 45 | 0.17 | 0.10 | 0.15 | 0.11 | 0.15 |
| 60 | 0.13 | 0.12 | 0.17 | 0.13 | 0.17 |
| 75 | 0.20 | 0.14 | 0.18 | 0.15 | 0.19 |
| 90 | 0.19 | 0.15 | 0.20 | 0.16 | 0.21 |
| 105 | 0.14 | 0.15 | 0.21 | 0.17 | 0.22 |
| 120 | 0.13 | 0.16 | 0.21 | 0.17 | 0.23 |
| 135 | 0.10 | 0.16 | 0.22 | 0.17 | 0.23 |

Note. — 68% and 90% confidence upper limits on $\epsilon_M$ are listed; i.e. $\epsilon_M^{68\circ}$ and $\epsilon_M^{90\circ}$ as described in §2.3.1. The geometric test assumes the X-rays are due to only hot gas in hydrostatic equilibrium with a potential generated by mass distributed like the optical light. In particular, rotation of the gas and emission from discrete sources are assumed negligible. The X-ray ellipticities ($\epsilon_X$) from Table 4 are listed to facilitate comparison.

Table 10: Geometric Test for Dark Matter - $\epsilon_M^{max}$ With Other Considerations

| $a$ | Rotation | | $F_{ds}/F_{hg} = 1/10$ | | $F_{ds}/F_{hg} = 1/3$ | | Rotation + Discrete | |
|---|---|---|---|---|---|---|---|---|
| (arcsec) | 68% | 90% | 68% | 90% | 68% | 90% | 68% | 90% |
| 30  | 0.10 | 0.13 | 0.09 | 0.12 | 0.08 | 0.11 | 0.08 | 0.11 |
| 45  | 0.12 | 0.16 | 0.10 | 0.14 | 0.10 | 0.14 | 0.11 | 0.14 |
| 60  | 0.14 | 0.19 | 0.13 | 0.17 | 0.13 | 0.17 | 0.14 | 0.18 |
| 75  | 0.17 | 0.21 | 0.14 | 0.19 | 0.16 | 0.20 | 0.18 | 0.22 |
| 90  | 0.18 | 0.23 | 0.16 | 0.20 | 0.18 | 0.23 | 0.20 | 0.25 |
| 105 | 0.19 | 0.24 | 0.16 | 0.22 | 0.19 | 0.24 | 0.22 | 0.26 |
| 120 | 0.19 | 0.24 | 0.17 | 0.22 | 0.19 | 0.25 | 0.23 | 0.27 |
| 135 | 0.19 | 0.25 | 0.17 | 0.23 | 0.20 | 0.26 | 0.23 | 0.29 |

Note. — Upper limits correspond to values of $\epsilon_M$ obtained from the geometric test for dark matter as in Table 9 except here the effects of rotation like that of the stars (§5.2.1), emission from discrete sources (§5.2.2), and a combination of both (using $F_{ds}/F_{hg} = 1/3$) are considered.

Table 11: Results for the Shape of the Composite Mass

| Mass Model | $\epsilon_{mass}$ 68% | $\epsilon_{mass}$ 90% | $\chi^2_{min}$ | $R_c$ (arcsec) | $|\Gamma|$ |
|---|---|---|---|---|---|
| Isothermal: | | | | | |
| SMD: $n = 1$, Oblate | 0.51 - 0.69 | 0.34 - 0.80 | 7.5 | 2.0 - 5.4 | 6.51 - 7.45 |
| SMD: $n = 1$, Prolate | 0.48 - 0.63 | 0.32 - 0.71 | 8.0 | 2.2 - 7.0 | 6.02 - 7.24 |
| SMD: $n = 1.5$, Oblate | 0.58 - 0.83 | 0.37 - 0.94 | 18.0 | 11.4 - 24.2 | 5.44 - 5.68 |
| SMD: $n = 1.5$, Prolate | 0.52 - 0.70 | 0.36 - 0.76 | 18.5 | 12.4 - 27.3 | 5.28 - 5.62 |
| SMD: Hernquist, Oblate | 0.52 - 0.72 | 0.33 - 0.83 | 8.0 | 71.9 - 105.2 | 5.74 - 6.23 |
| SMD: Hernquist, Prolate | 0.47 - 0.63 | 0.31 - 0.71 | 8.5 | 57.2 - 128.9 | 5.54 - 6.17 |
| SMD: $n = 1$, $i = 75\,\mathrm{deg}$, Oblate | 0.57 - 0.75 | 0.40 - 0.85 | 7.5 | 1.7 - 5.0 | 6.83 - 7.83 |
| SMD: $n = 1$, $i = 75\,\mathrm{deg}$, Prolate | 0.49 - 0.64 | 0.33 - 0.72 | 8.0 | 1.9 - 6.5 | 6.38 - 7.67 |
| SP: Logarithmic, Oblate | 0.57 - 0.73 | 0.38 - 0.82 | 8.3 | 4.4 - 8.9 | 5.69 - 6.12 |
| SP: Logarithmic, Prolate | 0.57 - 0.73 | 0.38 - 0.82 | 8.8 | 4.5 - 9.5 | 5.44 - 6.05 |
| SP: $n = 0.1$, Oblate | $\geq 0.67$ | $\geq 0.45$ | 10.1 | 8.3 - 13.7 | 9.65 - 10.73 |
| SP: $n = 0.1$, Prolate | $\geq 0.67$ | $\geq 0.45$ | 10.5 | 8.4 - 14.3 | 9.30 - 10.69 |
| Rotation | 0.36 - 0.51 | 0.22 - 0.60 | 9.5 | 50.1 - 86.9 | 5.78 - 6.26 |
| $F_{ds}/F_{hg} = 1/10$ | 0.49 - 0.69 | 0.31 - 0.80 | 7.8 | 2.4 - 6.7 | 6.36 - 7.27 |
| $F_{ds}/F_{hg} = 1/5$ | 0.47 - 0.69 | 0.26 - 0.80 | 7.8 | 3.0 - 9.5 | 6.25 - 6.98 |
| $F_{ds}/F_{hg} = 1/3$ | 0.41 - 0.68 | 0.21 - 0.80 | 8.0 | 4.4 - 19.4 | 5.95 - 6.58 |
| $F_{ds}/F_{hg} = 1/2$ | 0.35 - 0.62 | 0.16 - 0.80 | 11.0 | 8.8 - 52.3 | 5.64 - 6.96 |
| Rotation + Discrete | 0.26 - 0.45 | 0.12 - 0.56 | 7.6 | 80.2 - 160.0 | 5.32 - 5.82 |
| Polytropic: | | | | | |
| SMD: $n = 1$, Oblate | 0.51 - 0.69 | 0.33 - 0.80 | 7.0 | 0.10 - 12.3 | 3.70 - 11.9 |
| SMD: $n = 1$, Prolate | 0.47 - 0.63 | 0.30 - 0.71 | 7.1 | 0.12 - 11.9 | 3.48 - 11.1 |
| SMD: $n = 1.5$, Oblate | $\geq 0.75$ | $\geq 0.53$ | 7.3 | 1.0 - 12.0 | 2.14 - 4.0 |
| SMD: $n = 1.5$, Prolate | $\geq 0.63$ | $\geq 0.43$ | 7.4 | 1.2 - 12.0 | 2.14 - 4.0 |

Note. — This is a list of selected models for the composite mass. The listed rotation and discrete models are oblate. $\chi^2_{min}$ is the typical value in the 90% confidence interval (8 dof for isothermal models, 7 dof for polytropes). Similarly, $R_c$ and $\Gamma$ correspond to the 90% confidence limits. For models where only a lower limit for $\epsilon_{mass}$ is indicated, the ranges for $R_c$ and $\Gamma$ are estimates.

Table 12: Composite Mass and Mass-to-Light Ratio

| Mass Model | $M(h_{80}^{-1}10^{12}M_\odot)$ 68% | 90% | $\Upsilon_B(h_{80}\Upsilon_\odot)$ 68% | 90% |
|---|---|---|---|---|
| Isothermal: | | | | |
| SMD: $n = 1$, $a = 450''$, Oblate | 0.62 - 1.48 | 0.56 - 1.64 | 32.4 - 77.5 | 29.3 - 85.8 |
| SMD: $n = 1$, $a = 450''$, Prolate | 0.45 - 1.18 | 0.38 - 1.44 | 23.6 - 61.8 | 19.9 - 75.4 |
| SMD: $n = 1$, $a = min$, Oblate | 0.22 - 0.45 | 0.21 - 0.46 | 11.5 - 23.6 | 11.0 - 24.1 |
| SMD: $n = 1$, $a = min$, Prolate | 0.17 - 0.41 | 0.15 - 0.43 | 8.9 - 21.5 | 7.9 - 22.5 |
| SMD: Hernquist, $a = 450''$, Oblate | 0.32 - 0.78 | 0.31 - 0.82 | 16.7 - 40.8 | 16.2 - 42.9 |
| SMD: Hernquist, $a = 450''$, Prolate | 0.27 - 0.59 | 0.25 - 0.77 | 14.1 - 30.9 | 13.1 - 40.3 |
| SP: Logarithmic, $r = 450''$, Oblate | 0.78 - 1.71 | 0.75 - 1.75 | 40.8 - 89.5 | 39.2 - 91.6 |
| SP: Logarithmic, $r = 450''$, Prolate | 0.76 - 1.68 | 0.73 - 1.73 | 39.8 - 87.9 | 38.2 - 90.6 |
| (Hernquist) $F_{ds}/F_{hg} = 1/3$ | 0.44 - 1.04 | 0.41 - 1.12 | 23.0 - 54.4 | 21.5 - 58.6 |
| Rotation | 0.31 - 0.78 | 0.29 - 0.83 | 16.2 - 40.8 | 15.2 - 43.4 |
| Rotation + Discrete | 0.43 - 1.06 | 0.39 - 1.15 | 22.5 - 55.5 | 20.4 - 60.2 |
| Polytropic: | | | | |
| SMD: $n = 1$, $a = 450''$, Oblate | 0.28 - 2.84 | 0.19 - 3.18 | 14.6 - 148.6 | 9.9 - 166.4 |

Note. — Total integrated masses for selected models. For the SP models the mass is within a sphere of radius $r$. In all cases the 90% confidence limits on the temperature are used (see Table 7)

Table 13: Total Integrated X-ray Gas Mass

| $M_{gas}$ $(10^9 h_{80}^{-5/2} M_\odot)$ | $\bar{n}$ $(10^{-3} h_{80}^{1/2} \text{cm}^{-3})$ | $\bar{\tau}$ $(10^{10} h_{80}^{-1/2} \text{yr})$ |
|---|---|---|
| 0.69 - 1.81 | 2.70 - 7.11 | 0.90 - 1.25 |

Note. — Quantities represent 90% confidence limits within a sphere of radius $300''$ using the results for a single-temperature fit to the PSPC spectrum (Table 7).

Table 14: Dark Matter Shape Results

| $M_{DM}/M_{stars}$ | $\epsilon_{DM}$ | $\chi^2_{min}$ | $R_c$ (arcsec) | $|\Gamma|$ |
|---|---|---|---|---|
| 50 | 0.34 - 0.80 | 7.7 | 2.2 - 6.1 | 6.46 - 7.32 |
| 25 | 0.34 - 0.80 | 7.8 | 2.5 - 7.0 | 6.39 - 7.23 |
| 10 | 0.31 - 0.88 | 8.1 | 3.5 - 12.4 | 6.21 - 7.11 |
| 5 | $\geq 0.31$ | 8.5 | 7.7 - 76.5 | 6.10 - 7.11 |
| 4 | $\geq 0.31$ | 9.5 | 8.9 - 61.9 | 6.00 - 6.75 |
| 3 | $\geq 0.35$ | 16.3 | 9.2 - 55.2 | 5.90 - 6.37 |

Note. — 90% confidence limits on $\epsilon_{DM}$, $R_c$, and $\Gamma$ as a function of $M_{DM}/M_{stars}$ for the dark matter modeled as an oblate isothermal SMD with $n = 1$ and $a = 450''$. $\chi^2_{min}$ (8 dof) represents the typical minimum $\chi^2$ in the 90% interval. For the models having only a lower limit for $\epsilon_{DM}$ the 90% values for $R_c$ and $\Gamma$ are estimates.

---

Fig. 1.—
(a) Contour map of the X-ray surface brightness (0.4 - 2.4 keV) of the Lenticular galaxy NGC 1332 binned into $15''$ pixels for display; the contours are separated by a factor of 2 in intensity and the direction of Celestial North and East are indicated in the plot. The image has been corrected for the effects of exposure variations and telescopic vignetting. The image has been smoothed for visual clarity with a Gaussian of $\sigma = 15.0''$, although the image used for analysis is not smoothed in any manner. (b) Same as Figure (a) except the point sources listed in Table 2 have been removed by symmetric substitution (§2)

Fig. 2.—
Radial profile ($30''$ bins) corrected for the exposure variations, vignetting, and embedded point sources. The horizontal line is our estimate of the background level.

Fig. 3.—
The PSPC background-subtracted radial profile binned in $5''$ circular bins from $0'' - 15''$, $15''$ bins from $15'' - 105''$, then $45''$ bins to $285''$. The midpoint of each bin is use to define its location in the plot. Also shown are representations of the PSPC PSF for (1) $E = 0.818$ kev and $\theta = 7.25'$ (dashed), (2) $E = 1.1$ kev and $\theta = 7.25'$ (dotted), and (3) $E = 1.1$ kev and $\theta = 5'$ (solid).

Fig. 4.—
(a) The best-fit $\beta$-model (crosses) and the 90% confidence limits on $a_X$ and $\beta$ fit to the radial profile as in Figure 3. (b) Same as (a) except the inner $15''$ has been grouped into one bin.

Fig. 5.—
(a) The major-axis $I$-band profile of the combined MDM 1.3m and Tonry data. The solid line is the best-fit De Vaucouleurs Bulge + Exponential Disk model while the dotted line is the result of fitting a single De Vaucouleurs model. (b) The best-fit De Vaucouleurs Bulge + Exponential Disk model (crosses) of the $I$-band major-axis profile (Figure 5) convolved with the PSPC PSF and fitted to the PSPC radial profile prepared as in Figure 3.

Fig. 6.—
68% and 90% confidence contours for (a) Hydrogen column density vs. $T$ and (b) abundances vs. $T$ deduced from fitting a single-temperature Raymond-Smith plasma to the PSPC spectrum ($0'' - 120''$).

Fig. 7.—
Background-subtracted PSPC spectra for the circular inner region ($0'' - 30''$) and the annular outer region ($30'' - 120''$). The regions have comparable $S/N$.

Fig. 8.—
The dotted lines are the mean values of $\epsilon_M$ (plotted as perfect ellipses) for the X-ray surface brightness predicted from the geometric test for dark matter (i.e. mass follows light) assuming oblate symmetry; i.e. mean values of 1000 simulations. The solid lines are the observed $\epsilon_M$ (§2.3.1) for $a = 75'' - 90''$.

Fig. 9.—
(a) Radial profile of a typical isothermal SMD model (filled circles) consistent with the PSPC data (error bars). The model displayed is an oblate Hernquist SMD with $a = 450''$ and $\epsilon_{mass} = 0.60$. The best-fit parameters are $R_c = 72.8''$, $\Gamma = 5.95$, and $\chi^2_{min} = 10.0$. The 90% confidence contour and the best-fit parameter values are displayed in the inset. (b) X-ray isophotes for the best-fit model separated by a factor of 2 in intensity.

Fig. 10.—
(a) Radial profile of a typical isothermal SP model (filled circles) consistent with the PSPC data (error bars). The model displayed is an oblate Logarithmic SMD with $\epsilon_\Phi = 0.25$ which corresponds to $\epsilon_{mass} = 0.63$ using the aggregate shape determination explained in §6.2.2. The best-fit parameters are $R_c = 6.4''$, $\Gamma = 5.92$, and $\chi^2_{min} = 9.3$. The 90% confidence contour and the best-fit parameter values are displayed in the inset. (b) X-ray isophotes for the best-fit model separated by a factor of 2 in intensity.

Fig. 11.—
(a) Upper and lower solid (dashed) curves show 90% confidence limits of the integrated mass as a function of elliptical radius ($am$) for the isothermal SMD Hernquist model having $\epsilon_{mass} = 0.60$ and $a = 450''$. (b) same as (a) except $n = 1$ SMD model. (c) same as (a) except shown are the spherically averaged mass profiles of the isothermal logarithmic SP model with $\epsilon_\Phi = 0.25$ corresponding approximately to $\epsilon_{mass} = 0.63$.

Fig. 12.—
(a) Integrated mass as a function of elliptical radius ($am$) for the dark matter, stars, and gas. The dark matter is an $n = 1$ SMD, with $a = 450''$, $\epsilon_{mass} = 0.60$, and $M_{DM}/M_{stars} = 10$. Interior to $R_e \sim 50''$ (which is the radius enclosing half the light) $M_{DM}$ is comparable to $M_{stars}$. At larger radii, however, the dark matter dominates. The self-gravitation of the gas is not important anywhere in the galaxy. (b) Here we show the total mass for this model and the corresponding $\Upsilon_B$ as a function of ellipsoidal radius. $\Upsilon_B$ slowly rises interior to $R_e$ but increases dramatically at larger radii.

Fig. 13.—
(a) Shown are the 90% confidence limits on $\epsilon_M$ computed from the PSPC image (error bars; see Table 4) and the $\epsilon_M$ profiles computed from surface brightnesses of typical detailed isothermal models of the composite mass in §6.2. The mass models (all oblate) are (solid) $n = 1$ SMD with $\epsilon_{mass} = 0.60$ and $a = 450''$, (short dash - long dash) Hernquist SMD with $\epsilon_{mass} = 0.60$ and $a = 450''$, (dotted) Logarithmic SP with $\epsilon_\Phi = 0.25$ ($\epsilon_{mass} \sim 0.63$), (short dash) Hernquist SMD rotation model with $\epsilon_{mass} = 0.40$ and $a = 450''$, (long dash) $n = 1$ SMD with $\epsilon_{mass} = 0.60$ and $a = 450''$ and discrete component with $F_{ds}/F_{hg} = 1/3$, and (dot - short dash) Hernquist rotation model with $\epsilon_{mass} = 0.35$ and $a = 450''$ + discrete component with $F_{ds}/F_{hg} = 1/3$. (b) We show the true *isophotal* ellipticities of the X-ray surface brightness produced by the models of (a); by "true" we mean no PSF convolution; by "isophotal" we mean the ellipticity of an individual isophote, not of a large elliptical aperture.